\documentclass[article, shortnames,nojss]{jss}

\usepackage{orcidlink,thumbpdf,lmodern}

\usepackage{framed}
\usepackage{amsfonts}
\usepackage{xcolor}
\colorlet{red}{black}
\usepackage{amsmath}
\usepackage{algorithm}
\usepackage{algpseudocode}
\usepackage{subcaption}
\usepackage{booktabs}
\usepackage{enumerate}

\usepackage{cprotect}
\usepackage{comment}
\usepackage{subcaption}
\author{
Mary Lai O. Salva\~{n}a\orcidlink{0000-0003-4868-7713}%
\thanks{Mary Lai O. Salva\~{n}a and Sameh Abdulah contributed equally to this work.}\\
University of Connecticut\\
Storrs, CT, USA
\And
Sameh Abdulah\orcidlink{0000-0002-8850-5753}\footnotemark[1]\\
King Abdullah University\\
of Science and Technology\\
Thuwal, Saudi Arabia
\And
Minwoo Kim\orcidlink{0000-0003-4240-9878}\\
Pusan National University\\
Busan, South Korea
\AND
David Helmy\orcidlink{0009-0003-4709-5465}\\
BrightSkies Inc.\\
Alexandria, Egypt
\And
Ying Sun\orcidlink{0000-0001-6703-4270}\\
King Abdullah University\\
of Science and Technology\\
Thuwal, Saudi Arabia
\And
Marc G. Genton\orcidlink{0000-0001-6467-2998}\\
King Abdullah University\\
of Science and Technology\\
Thuwal, Saudi Arabia
}

\Plainauthor{Mary Salvana, Sameh Abdulah, Minwoo Kim, David Helmy, Ying Sun, Marc G. Genton}

\title{\pkg{MPCR}: Multi-Precision Computations Package in \proglang{R}}
\Plaintitle{\pkg{MPCR}: An R Package for Multi- and Mixed-Precision Computational Statistics}
\Shorttitle{\pkg{MPCR}: An \proglang{R} Package for Multi- and Mixed-Precision Computational Statistics}

\Abstract{
In the early days of computing, severe memory constraints made it necessary to use lower floating-point precision. As hardware capabilities have advanced, modern systems, particularly in computational statistics and scientific computing, have widely adopted 64-bit precision to reduce numerical errors and support complex calculations. However, in some applications, double-precision accuracy exceeds practical requirements, prompting interest in lower-precision alternatives that decrease computational complexity while maintaining adequate accuracy. This trend has accelerated with the advent of hardware optimized for low-precision computations, such as leveraging Tensor Cores technology in recent NVIDIA GPUs. Although lower precision can introduce numerical and accuracy challenges, many applications demonstrate robustness under these conditions. Consequently, new multi-precision algorithms have been developed to balance accuracy and computational cost. To facilitate the adoption of these approaches in statistical computing, this article introduces \pkg{MPCR}, a new \proglang{R} package that supports arithmetic operations at 16-, 32-, and 64-bit precision. Written in \proglang{C++} and integrated with \pkg{Rcpp}, \pkg{MPCR} delivers highly optimized multi-precision computations on both CPU and GPU, enabling seamless low-precision operations. Several examples demonstrate the benefits of \pkg{MPCR} across both performance and accuracy.
}


\Address{

  Mary Lai O. Salva\~{n}a\\
  University of Connecticut\\
 Department of Statistics\\
  Storrs, CT 06269, USA.\\
  E-mail: \email{marylai.salvana@uconn.edu}\\
  URL: \url{https://statistics.uconn.edu/person/mary-lai-salvana}

  Sameh Abdulah\\
  King Abdullah University of Science and Technology\\
  Extreme Computing Research Center (ECRC)\\
  Thuwal, 23955-6900, Saudi Arabia.\\
  E-mail: \email{sameh.abdulah@kaust.edu.sa}\\
  URL: \url{https://cemse.kaust.edu.sa/people/person/sameh-abdulah}

  Minwoo Kim\\
  Department of Statistics \\
  Pusan National University\\
  Busan, South Korea.\\
  E-mail: \email{mwkim@pusan.ac.kr}

  David Helmy\\
  BrightSkies Inc.\\
  Alexandria, Egypt.\\
  E-mail: \email{david.helmy@brightskiesinc.com}

  Ying Sun\\
 King Abdullah University of Science and Technology\\
  Extreme Computing Research Center (ECRC) 
   and
  Statistics Program\\
  Thuwal, 23955-6900, Saudi Arabia.\\
  E-mail: \email{ying.sun@kaust.edu.sa}\\
  URL: \url{https://cemse.kaust.edu.sa/es}
  
  Marc G. Genton\\
 King Abdullah University of Science and Technology\\
  Extreme Computing Research Center (ECRC) 
  and Statistics Program\\
  Thuwal, 23955-6900, Saudi Arabia\\
  E-mail: \email{marc.genton@kaust.edu.sa}\\
  URL: \url{https://cemse.kaust.edu.sa/stsds}
}

\begin{document}



\section[Introduction]{Introduction} \label{sec:intro}

Real numbers in computer architectures are encoded in bits and represented in floating-point format. Released in 1985, the \textit{IEEE 754-1985} standard defines floating-point formats and arithmetic \citep{kahan1996ieee, zuras2008ieee}. It underwent a revision in 2008, adding 16-bit half-precision alongside 128-bit quadruple-, 64-bit double-, and 32-bit single-precision formats. The original purpose of introducing the 16-bit half-precision format was for storage. Recently, it has proven helpful in scientific fields, including machine learning~\citep{bjorck2021low,tolliver2022comparative}, as well as linear algebra ~\citep{higham2019squeezing,higham2022mixed,scott2023algebraic,abdulah2024boosting,alomairy2025sustainably} and its various applications including ocean modeling \citep{tinto2019use}, differential equations \citep{ooi2020effect}, electromagnetic analysis \citep{masui2019research}, weather and climate modeling \citep{abdulah2024boosting}, and \textcolor{red}{molecular dynamics \citep{le2013spfp}, biology \citep{finn2011hmmer}, cosmology \citep{springel2005cosmological, habib2016hacc}, and plasma physics \citep{arber2018larexd, netti2023mixed}}. This trend has increased with the emergence of new hardware that can run low-precision computations substantially faster than full-precision, e.g., Graphics Processing Units (GPUs), Intel CPUs with DL boost, ARM CPUs, Tensor Processing Units (TPUs), Field-Programmable Gate Arrays (FPGAs), and AI hardware accelerators.
\textcolor{red}{Employing lower precision can potentially reduce computation times, and many applications demonstrate resilience to low-precision computations.} Although lowering precision can offer attractive acceleration, naively reducing it can lead to catastrophic round-off errors because the numerical representation range is limited.  This has prompted researchers to explore new lower- and multi-precision algorithms that balance accuracy and computational efficiency. {\color{red} In this paper, multi-precision algorithms refer to the use of different numerical precisions across distinct algorithmic stages, for example, performing most optimization iterations in single-precision and switching to double-precision for refinement.} 

Computational statistics is a field poised to benefit from these innovations~\citep{salvana2022parallel,cao2023reducing,abdulah2025high}. Recent work demonstrates that statistical algorithms can tolerate, and even exploit, reduced precision. For instance, \cite{guivant2023compressed} shows that storing the covariance matrix in lower precision can accelerate Kalman filtering by 50\%. \cite{misra2023vix} reported over $20\times$ speedups in variational inference using fixed-point arithmetic. \cite{maddox2022low} developed a mixed-precision conjugate gradient method for Gaussian process models, achieving up to $3\times$ speedup. Other examples include mixed-precision solutions for geostatistical Cholesky factorization \citep{abdulah2019geostatistical, cao2022reshaping, abdulah2022accelerating}. Such developments motivate the expansion of precision support in statistical software.

In \proglang{R} \citep{R2025A}, double-precision remains the default numeric type, while support for reduced precision is limited but can be enabled through extended packages. For instance, the \pkg{float} package enables single-precision computations when appropriate BLAS/LAPACK routines are available; however, extending this functionality to other formats, such as half-precision, presents significant challenges. For extended precision, packages such as \pkg{Rmpfr} \citep{Rmpfr} and \pkg{gmp} provide access to arbitrary-precision arithmetic via external libraries, but they are not designed for seamless integration with native \proglang{R} numeric types or high-performance linear algebra pipelines.

This paper introduces the \proglang{R} package \pkg{MPCR}, which provides multi-precision data structures and linear algebra computations within \proglang{R}. The package leverages \proglang{C++} and optimized BLAS/LAPACK backends to enable flexible precision selection for key numerical operations. \textcolor{red}{Specifically designed for researchers and data scientists working with multi-precision arithmetic, \pkg{MPCR} provides a convenient framework for efficient and accurate computations through well-designed data structures and carefully tuned operations on both CPU and GPU architectures.} To support diverse computational needs, the package allows users to create matrices and vectors in half-, single-, or double-precision formats on CPU and GPU and to combine these formats in multi-precision workflows. Examples from Markov Chain Monte Carlo, spatial statistics, principal component analysis, and Bayesian inference are also given to illustrate how \pkg{MPCR} can accelerate statistical analysis while maintaining numerical reliability. The rest of the paper is organized as follows: Section~\ref{sec:precision} discusses floating-point arithmetic and hardware support. Section~\ref{sec:MPCR} examines the internal design and workflow of the package. Section~\ref{sec:overview} provides an overview of the main functions, with examples and code snippets. Section~\ref{sec:performance} evaluates the performance of \pkg{MPCR} relative to native \proglang{R} across a range of linear algebra operations. Section~\ref{sec:applications} presents four statistical applications showcasing the package’s utility. We conclude in Section~\ref{sec:discuss}.

\section{Floating-Point Arithmetic in Modern Technology} \label{sec:precision}

The most widely used standard for floating-point arithmetic is \textit{IEEE 754-2019} (a revision of \textit{IEEE 754-2008}~\citep{kahan1996ieee, zuras2008ieee})~\citep{ieee7542019}. This standard significantly improved over its predecessor, \textit{IEEE 754-1985}, and includes a comprehensive set of guidelines that cover almost every aspect of floating-point (FP) theory, or, more simply, the approximation rules for real numbers on today's digital computers. The full name of this standard is \textit{IEEE Standard 754-2019 for Floating-Point Arithmetic}, which is often abbreviated as either \textit{IEEE 754-2019} or simply \textit{IEEE 754}. A floating-point number system is a finite subset $\mathcal{F}$ of $\mathbb{R}$, $\mathcal{F} \subset \mathbb{R}$, which depends on its elements $(\beta, t, e_{\text{min}}, e_{\text{max}})$. A number $x$ in $\mathcal{F}$ has the form:
\begin{align*}
    x = \pm m  \times \beta^{e-t+1},
\end{align*}
where $t$, $m$, and $e$ are integers known as \textit{precision}, \textit{significand} (\textit{mantissa}), and \textit{exponent}, respectively, and $\beta$ is the base, which is 2 for binary (commonly used on all current computers) and 10 for decimal. Here, the significand satisfies $0 \leq m \leq \beta^t -1$,  the exponent $e$ satisfies $e_{\text{min}} \leq e \leq e_{\text{max}}$, and the \textit{IEEE} standard requires that $e_{\text{min}} = 1 - e_{\text{max}}$. The number is stored in the computer in a format consisting of three fields: a sign bit, exponent bits, and significand bits. 

\begin{figure}[h!]
\centering
\includegraphics[width=0.7\textwidth]{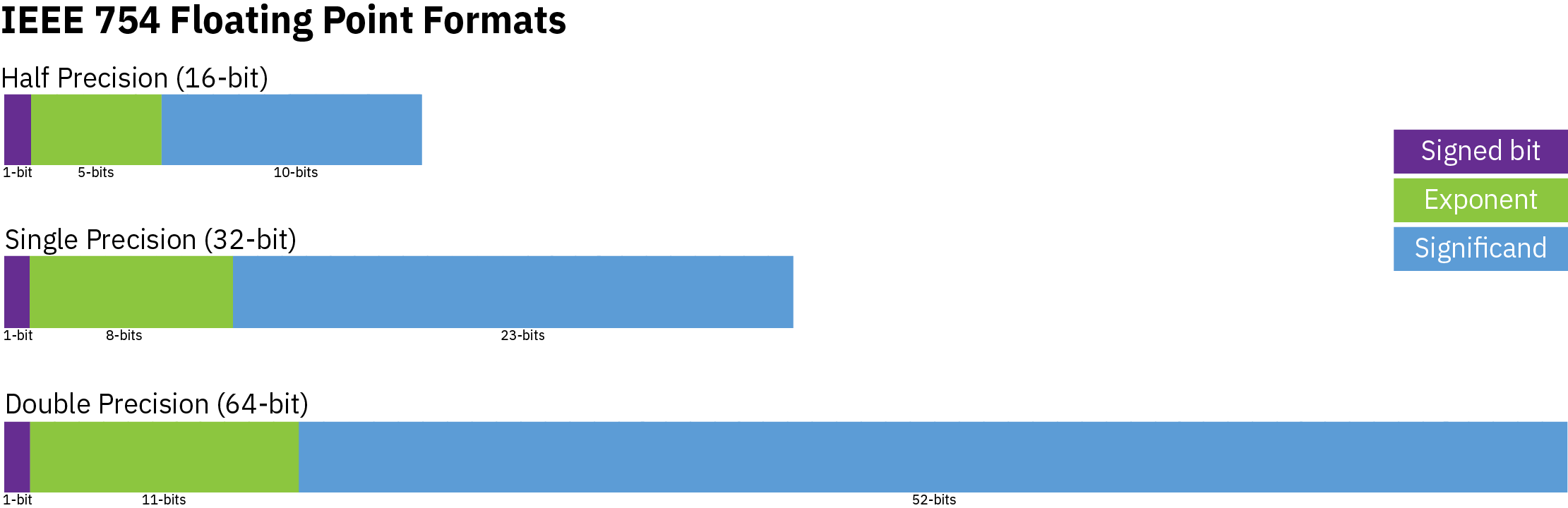}
\caption{The composition in bits of different precisions based on \textit{IEEE 754-2008} standards.}
\label{fig:ieee_754_floating_point_formats}
\end{figure}

Precision formats that take up more memory (in terms of bits) but can approximate real number values at very high accuracy are regarded as \textit{high} or \textit{full precision}, while those that consume smaller memory with lower accuracy are considered \textit{reduced} or \textit{low precision}. In the \textit{IEEE-754} standard, the double-precision 64-bit floating-point format, denoted FP64, is considered high precision, while the single-precision 32-bit floating-point format, denoted FP32, is regarded as low precision. FP32 uses 1-bit for the sign of the number, 8-bit for the exponent, and 23-bit for the significand and can represent values from $\pm (2 - 2^{-23}) \times 2^{127}$. On the other hand, FP64 uses 1-bit for the sign, 11-bit for the exponent, and 52-bit for the significand, and can represent values from $\pm (2 - 2^{-52}) \times 2^{1023}$. Another low-precision format is the half-precision 16-bit floating-point, denoted FP16. FP16 uses 1-bit for the sign, 5-bit for the exponent, and 10-bit for the significand and can represent values from $\pm (2 - 2^{-10}) \times 2^{15}$. Figure~\ref{fig:ieee_754_floating_point_formats} visualizes how a number~$x$ is represented in memory for each precision format.

The FP32 and FP64 are the most widely used formats as a broad range of off-the-shelf general-purpose processors natively support them. Although FP16 significantly reduces the memory requirements and increases the arithmetic throughput by $2\times$ against FP32 and $4\times$ against FP64, most commercial CPUs do not support FP16. However, recent chips from leading hardware vendors, including NVIDIA, Intel, and AMD, include FP16 arithmetic units to enhance performance in applications such as machine learning, gaming, and scientific computing. NVIDIA was among the first to support FP16, making it a storage format in CUDA 7.5 \citep{sabbagh2012low}. Since then, FP16 arithmetic has been supported on CUDA-enabled GPUs, beginning with the PASCAL architecture \citep{abdelfattah2021survey}. Other notable hardware supporting FP16 includes the NVIDIA P100 (2016) and V100 (2017) GPUs, as well as the A100 (2020), H100 (2022), and B100 (2024) GPUs, all of which feature Tensor Cores. One of the standout features of NVIDIA Tensor Cores is their ability to perform operations in mixed-precision, leveraging lower-precision (e.g., FP16) for specific calculations to improve performance and efficiency while still using higher-precision (e.g., FP32 and FP64) when higher accuracy is required. ARM processors also support 16-bit floating-point operations through dedicated floating-point and SIMD extensions (e.g., NEON and SVE). The instruction set reduces the memory footprint by reducing the code size. New ARM processors, such as the ARM Cortex series, support both 32-bit (ARM) and 16-bit (Thumb) instruction encodings to improve code density, while floating-point precision is provided by dedicated FP and SIMD units (e.g., NEON and SVE) \citep{higham2019simulating}. Additionally, there is growing anticipation that support for 8-bit FP operations may become more prevalent in the near future, reflecting the ongoing evolution in hardware and software to balance performance with computational efficiency.

\section{The MPCR Package Internal Design and Workflow}
\label{sec:MPCR}
For a long time, the \proglang{R} community has predominantly relied on 64-bit arithmetic for its computational tasks, seeking higher accuracy than is necessary. In practice, arithmetic with lower precision than 64-bit can offer faster execution and a smaller memory footprint while maintaining the same level of accuracy as 64-bit arithmetic in many applications. This work introduces the \pkg{MPCR} package, which supports lower-precision arithmetic. The \pkg{MPCR} package also provides low-precision arithmetic on GPUs, aiming to optimize computational efficiency without compromising accuracy in many applications.

\subsection{MPCR Package Internal Design} 

The \pkg{MPCR} package is designed with performance and flexibility in mind. We rely on \proglang{C++} to develop the package backend and port the code to \proglang{R} via the \pkg{Rcpp} package~\citep{eddelbuettel2011rcpp}. We use \proglang{C++} for various reasons: 1) \proglang{C++} excels in scenarios demanding high performance and efficiency; 2) \proglang{C++} is ideal for system-level programming that provides more flexibility in managing system resources, such as memory and processing power; 3) \proglang{C++} has many tools and libraries that can help build robust software; 4) \proglang{C++} can easily be integrated into different languages, including \proglang{R} and \proglang{Python}. 
Every function is invoked from the \proglang{R} environment, where \pkg{Rcpp} manages such calls, effectively bridging them to the corresponding \proglang{C++} functions in the \proglang{C++} layer.


The \pkg{MPCR} package offers basic operations and linear algebra operations. 
It supports numerous base functions in the \proglang{R} programming language, which can be readily applied to \pkg{MPCR}-objects. For instance, when performing the Cholesky factorization (\verb|chol()| in base \proglang{R}) on a single-precision \pkg{MPCR}-object, the \pkg{MPCR} package utilizes its internal \verb|chol()| function to execute in single-precision. Such a design facilitates integrating the \pkg{MPCR} package into any existing \proglang{R} code, enabling low-precision operations with minimal edits.

Furthermore, \pkg{MPCR} includes fully documented \proglang{R} and \proglang{C++} code, facilitating future development. The package was designed to be fully operational as a separate \proglang{C++} module to
be extended and used in any \proglang{C++} library. Additionally, \pkg{MPCR} uses template functions in \proglang{C++}, thereby avoiding code redundancy. Lastly, \pkg{MPCR} has a well-organized code structure, facilitating fast and easy code navigation.

\subsection{MPCR Package Workflow}
The process begins by specifying the appropriate precision for an \pkg{MPCR} object, which is then passed to the \proglang{C++} code via the \pkg{Rcpp} package. The {\bf precision controller} module in the \proglang{C++} level assigns the correct precision to the \pkg{MPCR} objects and their corresponding operations. The next step is to allocate memory on either CPU or GPU for the \pkg{MPCR} object, depending on its structure (e.g., vector or matrix) and precision (e.g., 32-bit for single-precision and 64-bit for double-precision). 
Once the object is allocated with particular precision, any call of a \proglang{C++} function for this object is managed by the \textit{dispatcher} module. This module dispatches operations to the appropriate template function based on the input precision. To link the \proglang{C++} function with the \proglang{R}  functions through the \pkg{MPCR} package, the \pkg{MPCR} \proglang{R} \textit{adapters} module is used. This module acts as a mapper between the \proglang{C++} and \proglang{R} environments. To provide a clearer description of each module, we provide examples of internal codes that illustrate its functionality.



\begin{Sinput}
### Code snippet for an MPCR R adapters module example at the C++ level 

The **MPCR R adapter** is a module that enables interaction between R and C++. 
It acts as a bridge between two incompatible interfaces.

std::vector <MPCR>
RSVD(MPCR *aInputA, const long &aNu, const long &aNv, const bool &aTranspose) 
{
    auto row = aInputA->GetNRow();
    auto col = aInputA->GetNCol();
    auto nv = aNv;
    auto nu = aNu;

    if (aNv < 0) {
        nv = std::min(row, col);
    }
    if (aNu < 0) {
        nu = std::min(row, col);
    }

    auto precision = aInputA->GetPrecision();
// Allocate three new objects for the SVD output.
    auto d = new MPCR(precision);
    auto u = new MPCR(precision);
    auto v = new MPCR(precision);

    SIMPLE_DISPATCH(precision, linear::SVD, *aInputA, *d, *u, *v, nu, nv,
                    aTranspose)

    std::vector <MPCR> output;
// Converting the output of C++ SVD funtion to a list, to match the shape of 
R Return values.
    output.push_back(*d);
    output.push_back(*u);
    output.push_back(*v);
    return output;
}

\end{Sinput}
\begin{Sinput}
### Code snippet for an MPCR dispatcher module example

**The Dispatcher** module main task is to choose the right signature for a 
function according to the input and output precision. This can be done with the 
help of the precision controller module.

**The Precision Controller** module main task is to decide the output precision
according to the input and the promotion strategy, and to decide the right
signature for the dispatcher.

MPCR* RRBind(MPCR *apInputA, MPCR *apInputB) 
{
// Getting the precision for object A and B.
    auto precision_a = apInputA->GetPrecision();
    auto precision_b = apInputB->GetPrecision();
// Using the precision controller to decide the output precision.
    auto output_precision = GetOutputPrecision(precision_a, precision_b);
// Allocating the output MPCR object with the right precision.
    auto pOutput = new MPCR(output_precision);
// Using the precision controller to decide the operation signature for 
the dispatcher to use.
    auto operation_comb = GetOperationPrecision(precision_a, precision_b,
                                                output_precision);
// Using the Operation signature to decide what template function to use.
    DISPATCHER(operation_comb, basic::RowBind, *apInputA, *apInputB, *pOutput)
    return pOutput;
}
\end{Sinput}

\subsection{Catch2}

\textit{Catch2} is a popular \proglang{C++} testing framework known for its comprehensive suite of tools that streamline the writing and maintenance of test cases for \proglang{C++} code. As a header-only library, it enables \proglang{C++} developers to easily create and manage their unit tests, offering a straightforward approach to testing. In the \pkg{MPCR} package, \textit{Catch2} provides many unit tests, facilitating thorough testing of the package's functionality. For example, integrating new features or enhancements into the package might unexpectedly impact existing code. Running these unit tests via \textit{Catch2} helps ensure software robustness throughout the development process. Furthermore, the \pkg{MPCR} package is designed to simplify the expansion of existing test cases, to support future development, and to maintain software reliability over time. Below is an example of using \textit{Catch2} in the \pkg{MPCR} package for the singular value decomposition function.


\begin{Sinput}
### Code snippets for C++ unit test for the SVD function using Catch2.

// Input Matrix Values.
        vector <double> values = {1, 1, 1, 1, 1, 1, 1, 1, 1, 1, 1, 1, 0, 0, 0,
                                  0, 0, 0, 0, 0, 0, 1, 1, 1, 0, 0, 0, 0, 0, 0,
                                  0, 0, 0, 1, 1, 1};
// Create an MPCR Object.
        MPCR a(values, FLOAT);
    
// Change MPCR Object to MPCR Matrix.
        a.ToMatrix(9, 4);
// Validate values for testing.
        vector <float> validate_values = {3.464102e+00, 1.732051e+00,
                                          1.732051e+00, 1.922963e-16};
// Allocating three MPCR objects for the SVD output
        MPCR d(FLOAT);
        MPCR u(FLOAT);
        MPCR v(FLOAT);
// This function will dispatch to SVD template function with the right precision.
        SIMPLE_DISPATCH(FLOAT, linear::SVD, a, d, u, v, a.GetNCol(),
                        a.GetNCol())
// Testing output d size.
        REQUIRE(d.GetSize() == 4);
        auto err = 0.001;
// Setting additional objects to test SVD Mathematically.
        MPCR dd(9, 4, FLOAT);
        for (auto i = 0; i < dd.GetSize(); i++) {
            dd.SetVal(i, 0);
        }
        for (auto i = 0; i < 4; i++) {
            dd.SetValMatrix(i, i, d.GetVal(i));
        }
        vector <double> temp_vals(81, 0);
        MPCR uu(temp_vals, FLOAT);
        uu.ToMatrix(9, 9);
        for (auto i = 0; i < u.GetSize(); i++) {
            uu.SetVal(i, u.GetVal(i));
        }

        MPCR vv = v;
        vv.Transpose();

// Performing  A = U Sigma V^H 
        MPCR temp_one(FLOAT);
        MPCR temp_two(FLOAT);

        SIMPLE_DISPATCH(FLOAT, linear::CrossProduct, uu, dd, temp_one, false,
                        false)

        SIMPLE_DISPATCH(FLOAT, linear::CrossProduct, temp_one, vv, temp_two,
                        false,
                        false)

        MPCR temp_three(FLOAT);
        SIMPLE_DISPATCH(FLOAT, math::Round, temp_two, temp_three, 1);

        SIMPLE_DISPATCH(FLOAT, math::PerformRoundOperation, temp_three,
                        temp_two, "abs");

// Checking if our SVD function values are valid mathematically
        for (auto i = 0; i < a.GetSize(); i++) {
            REQUIRE(temp_two.GetVal(i) == a.GetVal(i));
        }
\end{Sinput}

\section{Overview of the MPCR Package} \label{sec:overview}

To start using the \pkg{MPCR} package, it should be installed from the CRAN repository and imported into the \proglang{R}  environment by running these commands:
\begin{CodeChunk}
\begin{CodeInput}
R> install.packages("MPCR", type = "source")
R> library(MPCR)
\end{CodeInput}
\end{CodeChunk}

While CRAN hosts the stable \pkg{MPCR} package, our GitHub repository contains the latest development features and fixes. To download the package from the GitHub repository:

\begin{CodeChunk}
\begin{CodeInput}
R> remotes::install_github("https://github.com/stsds/MPCR")
\end{CodeInput}
\end{CodeChunk}

The \pkg{MPCR} package automatically enables or disables GPU support depending on whether CUDA is detected on the host system.

\subsection{Creating MPCR objects}
To create an \pkg{MPCR} object, i.e., matrix or vector, the constructor \verb|new| is invoked to specify the \verb|size| and \verb|precision| of the object. The default object created by the function \verb|new| is a zero-vector with a number of elements equal to the \verb|size| argument. To illustrate, the following codes create a single-precision \pkg{MPCR} zero-vector with 6 elements on CPU:
\begin{Schunk}
\begin{Sinput}
R> MPCR_object <- new(MPCR, 6, "single", "CPU")
R> MPCR_object
\end{Sinput}
\begin{Soutput}
MPCR Object: 32-Bit Precision on CPU
\end{Soutput}
\end{Schunk}

To create a half-precision \pkg{MPCR} zero-vector with 6 elements on GPU:
\begin{Schunk}
\begin{Sinput}
R> MPCR_object <- new(MPCR, 6, "half", "GPU")
R> MPCR_object
\end{Sinput}
\begin{Soutput}
MPCR Object: 16-Bit Precision on GPU
\end{Soutput}
\end{Schunk}

The function \verb|IsMatrix| can be used to check whether the newly created \verb|MPCR_object| is a vector or a matrix, e.g.:
\begin{Schunk}
\begin{Sinput}
R> MPCR_object$IsMatrix
\end{Sinput}
\begin{Soutput}
[1] FALSE
\end{Soutput}
\end{Schunk}

To display the values in \verb|MPCR_object|, use the \verb|PrintValues()| function:
\begin{Schunk}
\begin{Sinput}
R> MPCR_object$PrintValues()
\end{Sinput}
\begin{Soutput}
Vector Size : 6
---------------------
[ 1 ]       0      0      0      0      0      0
\end{Soutput}
\end{Schunk}

\verb|MPCR_object| can also be reconfigured as a matrix with a certain number of rows and columns. Suppose one requires \verb|MPCR_object| to be a $2 \times 3$ matrix. The following codes perform such transformation:
\begin{Schunk}
\begin{Sinput}
R> MPCR_object$ToMatrix(2,3)
R> MPCR_object$PrintValues()
\end{Sinput}
\begin{Soutput}
Precision  : 32-Bit  Precision 
Number of Rows : 2
Number of Columns : 3
---------------------
 [    0    0    0    ]
 [    0    0    0    ] 
\end{Soutput}
\end{Schunk}

Elements inside any \pkg{MPCR} object can be extracted and their values replaced by indicating the appropriate indices, e.g.: 
\begin{Schunk}
\begin{Sinput}
R> MPCR_object[1,1] <- 1
R> MPCR_object[1,2] <- 2
R> MPCR_object$PrintValues()
\end{Sinput}
\begin{Soutput}
Precision  : 32-Bit  Precision 
Number of Rows : 2
Number of Columns : 3
---------------------
 [    1    2    0    ]
 [    0    0    0    ]  
\end{Soutput}
\end{Schunk}

The \pkg{MPCR} objects can also be converted to \proglang{R} objects by invoking the \pkg{MPCR} function \verb|MPCR.ToNumericVector()| as follows:
\begin{Schunk}
\begin{Sinput}
R> matrix(MPCR.ToNumericVector(MPCR_object), nrow = 2, ncol = 3)
\end{Sinput}
\begin{Soutput}
     [,1] [,2] [,3]
[1,]    1    2    0
[2,]    0    0    0
\end{Soutput}
\end{Schunk}

Another approach to constructing an \pkg{MPCR} object is to convert the usual \proglang{R} object to \pkg{MPCR} objects using the function \verb|as.MPCR()|. In the following, we create a $3 \times 3$ \proglang{R} matrix and convert it into a single-precision $3 \times 3$ \pkg{MPCR} matrix:
\begin{Schunk}
\begin{Sinput}
R> a <- matrix(1:9, 3, 3)
R> MPCR_matrix <- as.MPCR(a,nrow=3,ncol=3,precision="single")
R> MPCR_matrix$PrintValues()
\end{Sinput}
\begin{Soutput}
Precision  : 32-Bit  Precision 
Number of Rows : 3
Number of Columns : 3
---------------------
 [    1    4     7    ]
 [    2    5     8    ]
 [    3    6     9    ]
\end{Soutput}
\end{Schunk}

\subsection{Operations on MPCR objects}
The \pkg{MPCR} package enables multi-precision arithmetic. Arithmetic operations on \pkg{MPCR} objects can be carried out using the same symbols used to perform those operations on \proglang{R} objects. With multi-precision arithmetic, one can add or subtract matrices with different precisions such that the resulting \pkg{MPCR} object inherits the precision of the \pkg{MPCR} object with the highest precision. In the following, a double-precision $2 \times 10$ matrix is added to a single-precision $2 \times 10$ matrix, resulting in a double-precision $2 \times 10$ matrix.
\begin{Schunk}
\begin{Sinput}
R> s1 <- as.MPCR(1:20, nrow=2, ncol=10, "single")
R> s2 <- as.MPCR(21:40, nrow=2, ncol=10, "double")
R> x <- s1 + s2
R> typeof(x)
R> x$PrintValues()
\end{Sinput}
\begin{Soutput}
MPCR Object : 64-Bit Precision
\end{Soutput}
\end{Schunk}

Other usual operations on \proglang{R} vectors and matrices can also be done on \pkg{MPCR} vectors and matrices by invoking the conventional function call in \proglang{R}. For example, the transpose of \verb|MPCR_matrix| in the previous example can be obtained by \verb|t(MPCR_matrix)| and multiplying two \pkg{MPCR} matrices is done using the product operator \verb|%*%|, e.g., \verb|s1 %*% t(s2)|. 

Linear algebra operations are also available and are called as they are in the base \proglang{R} package, such as \verb|chol()|, \verb|chol2inv()|, \verb|solve()|, and \verb|eigen()|. Furthermore, common functions used to evaluate matrices/vectors in \proglang{R} are also made available for \pkg{MPCR} objects such as \verb|cbind|, \verb|rbind|, \verb|diag|, \verb|min|, \verb|max|, etc., and can be invoked the same way as the base \proglang{R} package.

\section{Linear Algebra Functions Performance} \label{sec:performance}

This section presents a performance evaluation of key linear algebra operations in half-, single-\, and double-precision, implemented with the \pkg{MPCR} package and executed on both CPU and GPU platforms. 
All experiments in this and the following section were conducted on a dual-socket Intel Cascade Lake system (20 cores per CPU) with an NVIDIA V100 (32 GB) GPU, using CUDA 12.2 and Intel MKL. Our analysis focuses on computationally intensive kernels and on operations that require algorithmic or implementation-level modifications to achieve higher performance. Because cuBLAS provides the most mature and optimized half-precision support for matrix–matrix multiplication, we restrict our half-precision experiments to this operation.

\begin{figure*}[t!]
    \centering
    \begin{subfigure}[t]{0.32\textwidth}
        \centering
        \includegraphics[width=\linewidth]{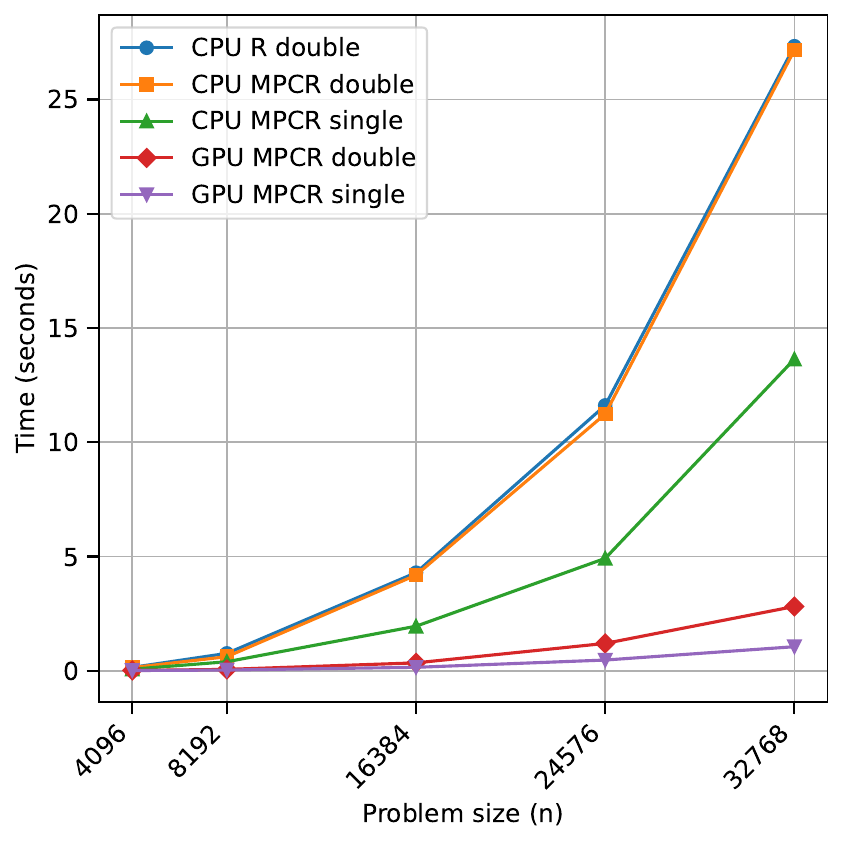}
        \caption{Cholesky factorization (\texttt{chol})}
        \label{fig:perf1_double}
    \end{subfigure}
    \hfill
    \begin{subfigure}[t]{0.32\textwidth}
        \centering
        \includegraphics[width=\linewidth]{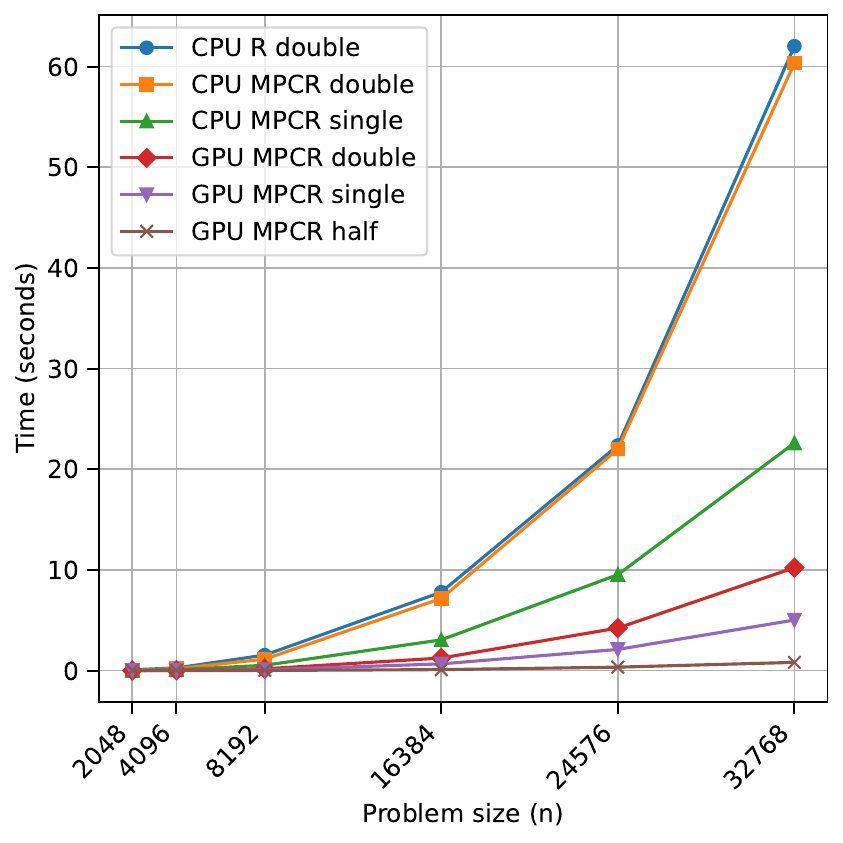}
        \caption{Matrix multiplication (\texttt{crossprod})}
        \label{fig:perf1_single}
    \end{subfigure}
    \hfill
    \begin{subfigure}[t]{0.32\textwidth}
        \centering
        \includegraphics[width=\linewidth]{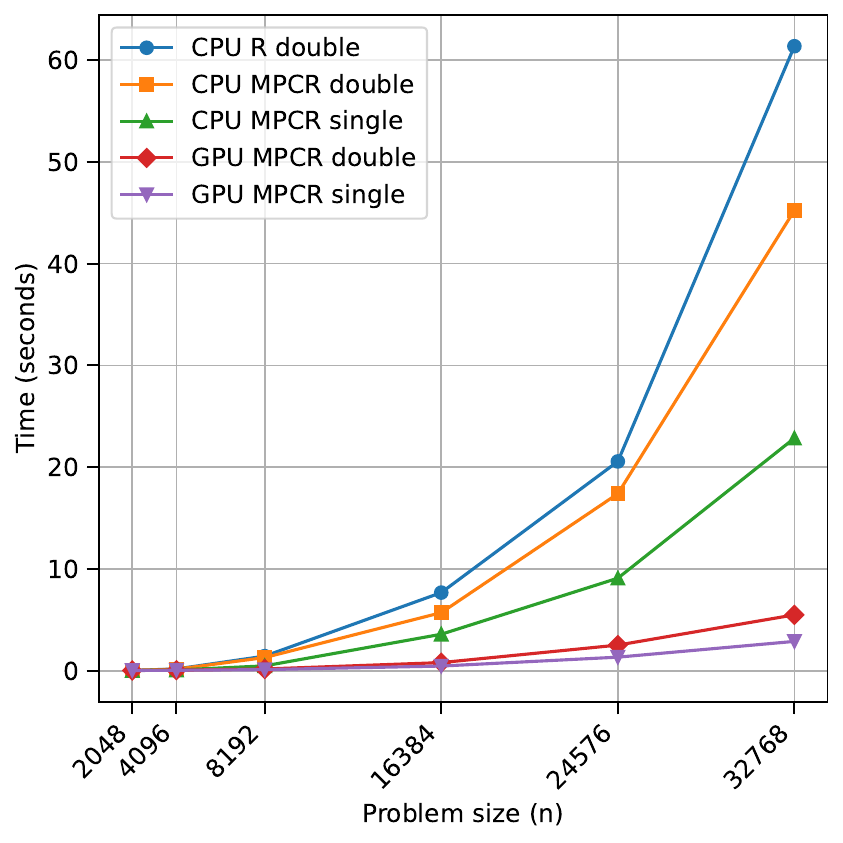}
        \caption{Triangular back substitution (\texttt{backsolve})}
        \label{fig:perf1_half}
    \end{subfigure}
        \cprotect\caption{Total execution time of linear algebra operations using the \pkg{MPCR} package compared with double-precision \proglang{R} baselines across CPU and GPU platforms.
(a) Cholesky factorization (\verb|chol()|), (b) matrix multiplication (\verb|crossprod()|), and (c) triangular back substitution (\verb|backsolve()|). Results are reported for double- and single-precision MPCR implementations alongside CPU \proglang{R} double-precision.}
    \label{fig:perf1}
\end{figure*}

Figure~\ref{fig:perf1} illustrates the performance of \pkg{MPCR} double- and single-precision operations relative to \proglang{R} double-precision, with half-precision reported only for matrix multiplication (crossprod). Both \proglang{R} and \pkg{MPCR} are linked against \emph{Intel MKL} to ensure a fair comparison across all evaluated operations. 

Figure~\ref{fig:perf1_double} shows the performance of the Cholesky factorization computed using five configurations: (i) double-precision with the native \proglang{R} implementation on CPU, (ii) double- and single-precision executions using \pkg{MPCR} on CPU, and (iii) double- and single-precision executions using \pkg{MPCR} on GPU. 
The figure shows that native \proglang{R} and CPU-based \pkg{MPCR} in double-precision have similar runtimes, with \pkg{MPCR} providing modest but consistent improvements (up to $1.24\times$). This reflects reduced overhead in \pkg{MPCR}, which performs most computations in optimized \texttt{C++} and calls BLAS routines directly. Using single-precision on CPU further lowers execution time by roughly $50\%$ at the largest size (from $27.16\mathrm{s}\ \mathrm{to}\ 13.64\mathrm{s}$ at $n=32768$), corresponding to about a $2.0\times$ speedup over CPU double-precision.

Offloading to the GPU yields substantially larger gains. At $n=32768$, GPU double-precision requires $2.81\mathrm{s}$, achieving a $9.7\times$ speedup over native \proglang{R}$\,(27.33\mathrm{s})$ and a $9.6\times$ speedup over CPU double-precision \pkg{MPCR}. GPU single-precision further reduces runtime to $1.05\mathrm{s}$, corresponding to a $26.0\times$ speedup over \proglang{R}. Relative Frobenius errors remain small, at $1.501608\times 10^{-4}$ for single-precision and $2.60685\times 10^{-13}$ for double-precision.

In Figure~\ref{fig:perf1_single}, for matrix–matrix multiplication (\verb|crossprod()|), GPU-based double-precision \pkg{MPCR} reduces the runtime from $62.02$ seconds (CPU \proglang{R} double) to $10.22$ seconds at the largest problem size ($n=32768$), corresponding to a $6.1\times$ speedup. GPU-based single-precision further improves performance to $5.02$ seconds, yielding a $12.4\times$ speedup over \proglang{R}. These gains are largely driven by highly optimized cuBLAS kernels and an efficient mapping of \verb|crossprod()| to GPU hardware. GPU-based half-precision achieves $0.811$ seconds at $n=32768$, outperforming CPU \proglang{R} double by $76.5\times$ and CPU \pkg{MPCR} double by $74.4\times$; relative Frobenius errors are $3.501115\times 10^{-3}$ at $n=4096$ for half-precision, $2.173175\times 10^{-7}$ for single-precision, and $1.581076\times 10^{-15}$ for double-precision.

In Figure~\ref{fig:perf1_half}, a similar trend is observed for the triangular solve (backsolve). GPU-based double-precision \pkg{MPCR} reduces the runtime from $61.36\mathrm{s}$ (CPU \proglang{R} double) to $5.49\mathrm{s}$ at $n=32{,}768$, corresponding to an $11.2\times$ speedup. GPU-based single-precision further improves performance to $2.88\,\mathrm{s}$, achieving up to a $21.3\times$ speedup over native \proglang{R}. These results highlight the efficiency of the GPU implementation and the benefit of reduced precision for compute-intensive triangular solves.

Below is an example of the crossproduct operation using five different computation variants.

\begin{Schunk}
\begin{Sinput}
R> n <- 4096

R> A0 <- matrix(runif(n*n), n, n)
R> B0 <- matrix(runif(n*n), n, n)

R> MPCR.SetOperationPlacement("CPU")
R> A <- as.MPCR(A0, n, n, precision="double", placement="CPU")
R> B <- as.MPCR(B0, n, n, precision="double", placement="CPU")
R> C_double_cpu <- A %*% B

R> A <- as.MPCR(A0, n, n, precision="single", placement="CPU")
R> B <- as.MPCR(B0, n, n, precision="single", placement="CPU")
R> C_single_cpu <- A %*% B

R> MPCR.SetOperationPlacement("GPU")
R> A <- as.MPCR(A0, n, n, precision="double", placement="GPU")
R> B <- as.MPCR(B0, n, n, precision="double", placement="GPU")
R> C_double_gpu <- A %*% B

R> A <- as.MPCR(A0, n, n, precision="single", placement="GPU")
R> B <- as.MPCR(B0, n, n, precision="single", placement="GPU")
R> C_single_gpu <- A %*% B

R> A <- as.MPCR(A0, n, n, precision="half", placement="GPU")
R> B <- as.MPCR(B0, n, n, precision="half", placement="GPU")
R> C_half_gpu <- A %*% B

R> C_ref <- A0 %*% B0
R> rel_err_single <- norm(as.matrix(C_single_gpu) - C_ref, "F") /
+                    norm(C_ref, "F")
\end{Sinput}
\end{Schunk}

\section{Applications} \label{sec:applications}

This section presents examples demonstrating how \pkg{MPCR} improves the efficiency of computationally intensive Frequentist and Bayesian workflows. \pkg{MPCR} integrates seamlessly with existing \proglang{R} code, enabling users to select the appropriate arithmetic precision based on accuracy and performance requirements. Importantly, the objective is not to enforce lower precision, which may compromise accuracy, but to provide a flexible and systematic framework for exploiting multiple arithmetic precisions when advantageous.

\subsection{Metropolis-Hastings Algorithm in High Dimensions}

The Metropolis-Adjusted Langevin Algorithm (MALA) is a gradient-informed variant of the Metropolis–Hastings method designed for sampling in high-dimensional settings \citep{metropolis1953equation,hastings1970monte}. Its computational cost is dominated by repeated linear algebra operations, making the choice of precision particularly impactful.

We sample from a 2D spatial Gaussian random field defined on an $M\times M$ grid with target distribution $\mathcal{N}_{n}(\boldsymbol{\mu}, \boldsymbol{\Sigma})$, where $n=M^2$. MALA uses the proposal
\[
q(\mathbf{z}|\mathbf{z}^{\text{current}})=
\mathcal{N}_{n}\!\left(
\mathbf{z}^{\text{current}}+\frac{h}{2}\mathbf{M}\nabla\log p(\mathbf{z}^{\text{current}}),
\, h\mathbf{M}
\right),
\]
requiring repeated matrix–vector products and linear solves.

We compare five different settings: standard \proglang{R} double-precision, \pkg{MPCR}-Double, and \pkg{MPCR}-Single executed on both CPU and GPU. The implementation is shown in the code below.
\begin{Schunk}
\begin{Sinput}
R> M <- 120; n <- M^2 # Setup: spatial grid
R> locs <- expand.grid(x=(0:(M-1))/(M-1), y=(0:(M-1))/(M-1))
R> D <- as.matrix(dist(locs))
R> 
R> # Target distribution: mean=0, covariance=exp(-D/0.5)
R> mu <- rep(0, n); sig <- exp(-D/0.5)
R> # Preconditioning matrix for MALA
R> pre_M <- exp(-D/0.05); h <- 0.01; I <- 100
\end{Sinput}
\end{Schunk}
\begin{Schunk}
\begin{Sinput}
R> # Function to run MALA with different precisions
R> run_mala <- function(prec='R-Double') {
+   # Convert to MPCR if needed (consistent precision throughout)
+   use_mpcr <- grepl('MPCR', prec)
+   if(use_mpcr) {
+     # Precision
+     p <- if (prec == "MPCR-Single-CPU" || prec == "MPCR-Single-GPU") 
+           "single" else "double"
+     # Hardware placement to expected values
+     op_place  <- if (prec == "MPCR-Single-CPU" || prec == "MPCR-Double-CPU")
+                  "CPU" else "GPU"
+     mu <- as.MPCR(mu, n, 1, p, placement = op_place)
+     sig <- as.MPCR(sig, n, n, p, placement = op_place)
+     pre_M <- as.MPCR(pre_M, n, n, p, placement = op_place)
+     MPCR.SetOperationPlacement(placement = op_place)
+   }
+   
+   # Precompute matrices
+   sig.inv <- solve(sig)
+   L <- t(chol(if(use_mpcr) pre_M$PerformMult(h) else h*pre_M))
+   pre_M_inv <- solve(if(use_mpcr) pre_M$PerformMult(h) else h*pre_M)
+   
+   # Gradient and gradient step
+   grad <- function(x) sig.inv %*% x
+   step <- function(x, t) {
+     if(use_mpcr) x - (pre_M %*% grad(x))$PerformMult(t)
+     else x - t * pre_M %*% grad(x)
+   }
+   
+   # Initialize chain
+   set.seed(1234); z <- runif(n)
+   if(use_mpcr) z <- as.MPCR(z, n, 1, p)  # Use consistent precision
+   trace <- matrix(NA, n, I)
+   
+   # MALA iterations
+   t0 <- Sys.time()
+   for(i in 1:I) {
+     set.seed(i)
+     eps <- if(use_mpcr) as.MPCR(rnorm(n), n, 1, p) else rnorm(n)
+     z_prop <- step(z, 0.5*h) + L %*% eps
+     
+     # Compute log ratio (helper for MPCR conversion)
+     num <- function(x) if(use_mpcr) MPCR.ToNumericVector(x) else as.numeric(x)
+     p_prop <- -0.5 * num(t(z_prop-mu) %*% sig.inv %*% (z_prop-mu))
+     p_curr <- -0.5 * num(t(z-mu) %*% sig.inv %*% (z-mu))
+     q_curr <- -0.5 * num(t(z-step(z_prop,0.5*h)) %*% pre_M_inv %*% 
+                          (z-step(z_prop,0.5*h)))
+     q_prop <- -0.5 * num(t(z_prop-step(z,0.5*h)) %*% pre_M_inv %*% 
+                          (z_prop-step(z,0.5*h)))
+     
+     if(runif(1) < exp(min(0, p_prop - p_curr + q_curr - q_prop)))
+       z <- z_prop
+     trace[,i] <- if(use_mpcr) MPCR.ToNumericVector(z) else z
+   }
+   list(trace=trace, time=difftime(Sys.time(), t0, units="mins"))
+ }
\end{Sinput}
\end{Schunk}
\begin{Schunk}
\begin{Sinput}
R> # Run all precisions and create figure
R> res <- lapply(c('R-Double','MPCR-Double-CPU','MPCR-Single-CPU', 
+         'MPCR-Double-GPU','MPCR-Single-GPU'), run_mala)
R> names(res) <- c('R-Double','MPCR-Double-CPU','MPCR-Single-CPU', 
+         'MPCR-Double-GPU','MPCR-Single-GPU')
R> 
R> # Visualization 
R> pdf("mh_results.pdf", width=16, height=4)
R> par(mfrow=c(1,4), mar=c(5,5,3,4)) 
R> cols <- rev(rainbow(100, start=0, end=4/6)) 
R> # Panel 1: Initial z0
R> set.seed(1234); z0 <- matrix(runif(n), M, M)
R> image(z0, main=expression(bold(z[0])), col=cols, cex.lab=1.5,
+       xlab=expression(s[x]), ylab=expression(s[y]), axes=FALSE)
R> axis(1); axis(2); box()
R> # Add legend using fields package
R> library(fields)  # For image.plot with legend
R> image.plot(z0, col=cols, legend.only=TRUE, add=TRUE)
R> # Panels 2-4: Final states with timing
R> for(nm in names(res)) {
+   zf <- matrix(res[[nm]]$trace[,I], M, M)
+   image(zf, main=sprintf("%s (%.2f mins)", nm, res[[nm]]$time), cex.lab=1.5,
+         col=cols, xlab=expression(s[x]), ylab=expression(s[y]), axes=FALSE)
+   axis(1); axis(2); box()
+   image.plot(zf, col=cols, legend.only=TRUE, add=TRUE)
+ }
R> dev.off()
\end{Sinput}
\end{Schunk}

\begin{figure*}[t!]
    \centering
    \includegraphics[width=\textwidth]{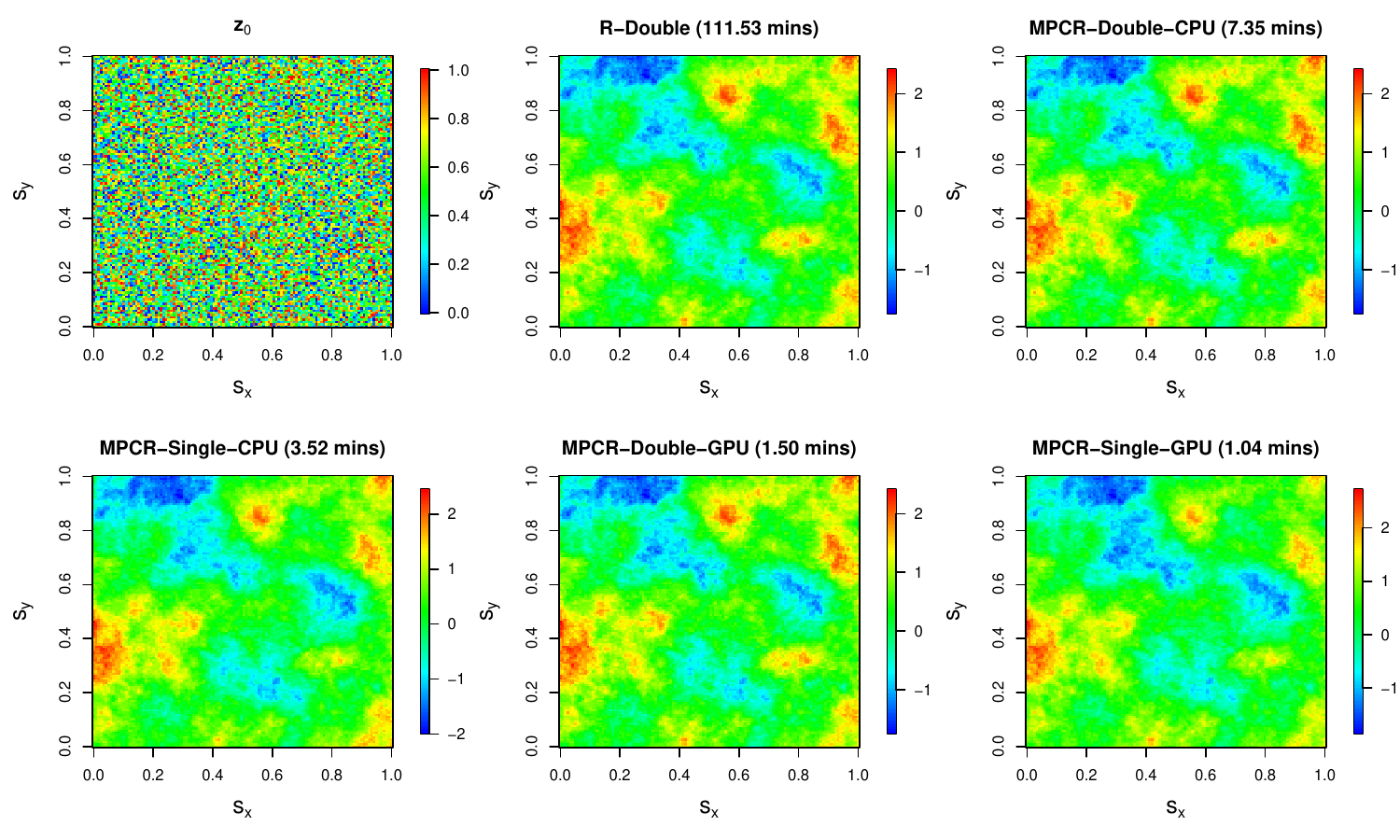}
        \caption{Initial field $z_0$ and MALA output after 2000 iterations ($M=120$). All precisions produce visually similar fields; only the computation time differs.}
    \label{fig:mh_results}
\end{figure*}
\begin{figure}[t!]
\centering
\includegraphics[width=0.7\textwidth]{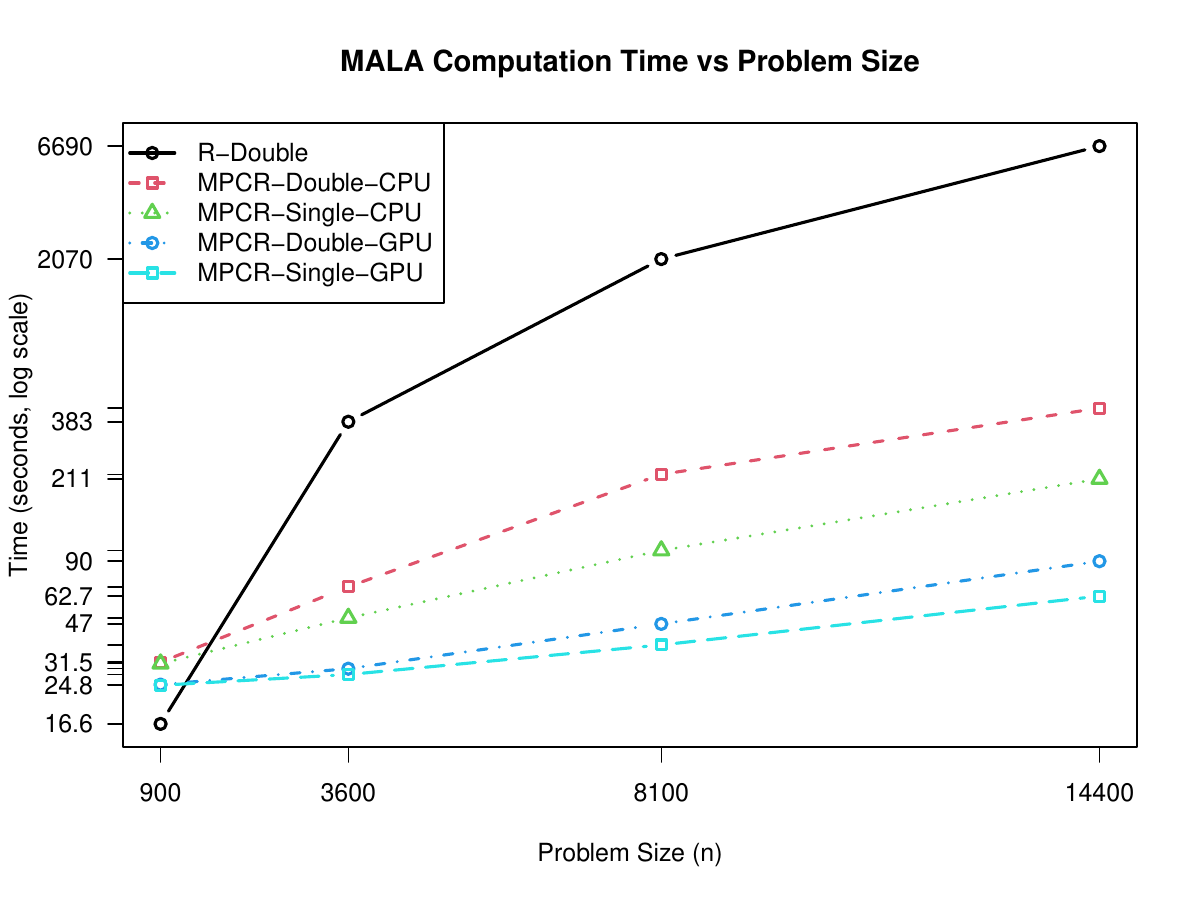}
\caption{MALA computation time (seconds) across problem sizes (up to $14{,}000$).}
\label{fig:MPCR_JSS_application_mh_result_time}
\end{figure}

Figure~\ref{fig:mh_results} shows the spatial fields obtained after 2000 iterations for each precision setting. Figure~\ref{fig:MPCR_JSS_application_mh_result_time} shows the timing results across problem sizes. \pkg{MPCR}-Single on GPU delivers the best overall performance, followed by \pkg{MPCR}-Double on GPU, then \pkg{MPCR}-Single on CPU, while \pkg{MPCR}-Double on CPU still provides a substantial improvement over standard \proglang{R}.  Across all five settings, the methods produce nearly identical realizations, while runtimes differ markedly. For n=$14{,}400$, Standard \proglang{R} with double-precision is consistently the slowest, requiring {\color{red}111.53} minutes in a representative case. \pkg{MPCR} with double-precision on CPU substantially reduces runtime to {\color{red}7.35} minutes, followed by \pkg{MPCR} with single-precision on CPU ({\color{red}3.52} minutes). GPU acceleration yields further gains, with \pkg{MPCR}-Double on GPU outperforming its CPU counterpart, and \pkg{MPCR}-Double on GPU achieves an overall time of {\color{red}1.5} minutes, while \pkg{MPCR}-Single on GPU achieves the best overall performance at {\color{red}1.04} minutes. In this example, single-precision yields the largest speedup, but the broader capability of flexible precision selection is the critical advantage.

\subsection{Maximum Likelihood Estimation of Mat\'ern Covariance Function}

\begin{figure}[t!]
\centering
\includegraphics[width=7cm]{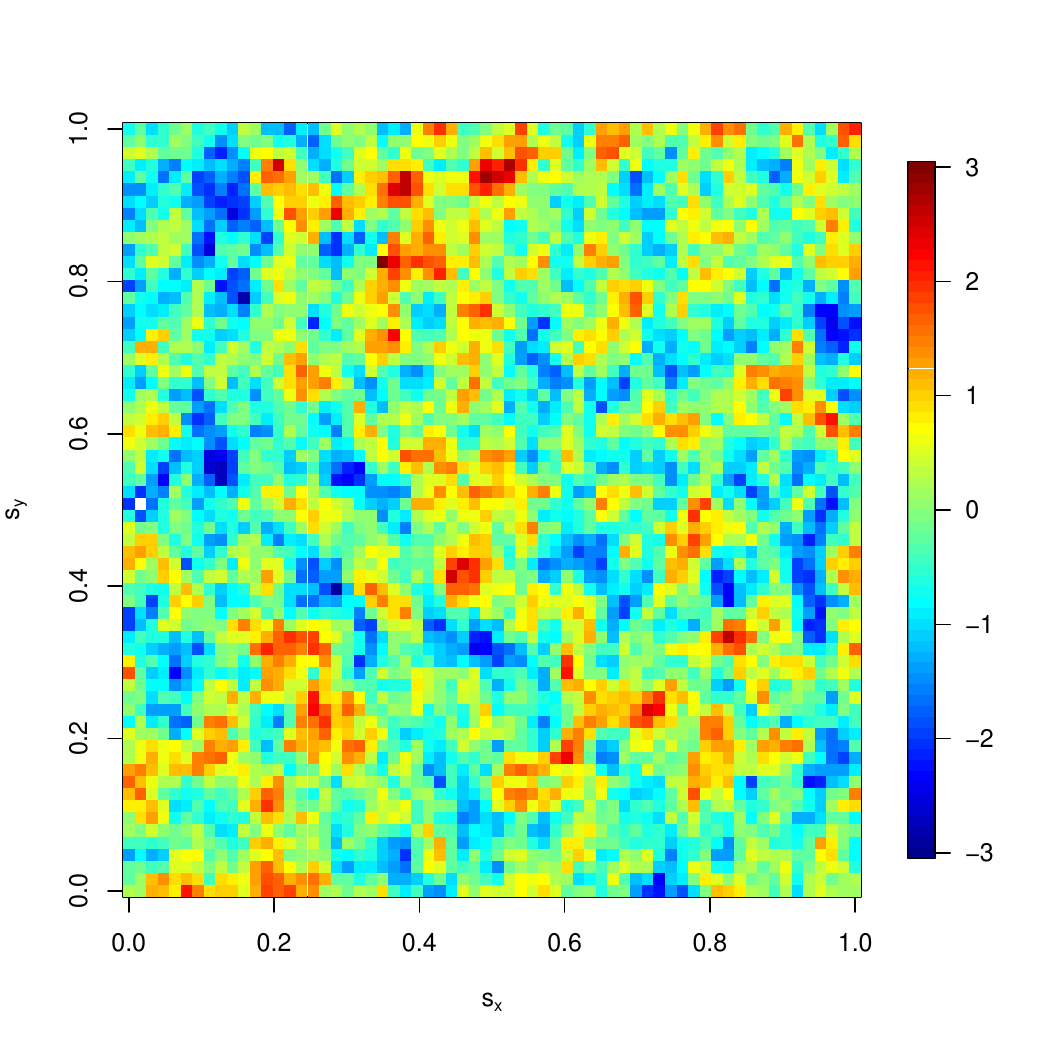}
\caption{Sample spatial field on a 100 $\times$ 100 grid in a unit square generated from a Mat\'{e}rn spatial covariance function model with $\boldsymbol{\theta} = (0.03, 0.5, 1)^\top$.}
\label{fig:MPCR_JSS_application_mle_simulated_values}
\end{figure}

Our second example illustrates how \pkg{MPCR} can accelerate likelihood-based inference in spatial statistics. Consider a Gaussian random field observed at $n=14{,}400$ locations on a $120\times 120$ grid (Figure~\ref{fig:MPCR_JSS_application_mle_simulated_values}). The field is modeled using the Matérn covariance function
\[
\mathrm{cov}\{Z(\mathbf{s}_i),Z(\mathbf{s}_j)\}
= 
\frac{\sigma^2}{2^{\nu-1}\Gamma(\nu)}
\mathcal{M}_\nu\!\left(\frac{\|\mathbf{s}_i-\mathbf{s}_j\|}{a}\right),
\]
with parameters $\boldsymbol{\theta}=(\nu,a,\sigma^2)^\top$ \citep{guttorp2006studies,matern2013spatial,wang2023parameterization}. Parameter estimation is carried out via the Gaussian log-likelihood
\[
\ell(\boldsymbol{\theta})
= -\frac{n}{2}\log(2\pi)
   -\frac{1}{2}\log|\boldsymbol{\Sigma}(\boldsymbol{\theta})|
   -\frac{1}{2}\mathbf{Z}^{\top}\boldsymbol{\Sigma}(\boldsymbol{\theta})^{-1}\mathbf{Z}.
\]

Evaluating $\ell(\boldsymbol{\theta})$ is computationally demanding, as storing the $n\times n$ covariance matrix requires $O(n^2)$ memory while computing its Cholesky factorization incurs $O(n^3)$ operations. These costs are incurred at every optimizer iteration, making the precision choice highly relevant to performance.

To evaluate \pkg{MPCR}, we construct $\boldsymbol{\Sigma}(\boldsymbol{\theta})$ using five formats: standard \proglang{R} double-precision, \pkg{MPCR} double-precision on CPU, \pkg{MPCR} single-precision on CPU, \pkg{MPCR} double-precision on GPU, and \pkg{MPCR} single-precision on GPU. We assume an exponential covariance function (i.e., a Matérn covariance with $\nu = 0.5$), $C(h) = \sigma^{2} \exp\!\left(-\frac{\|\mathbf{s}_i-\mathbf{s}_j\|}{a}\right)$. This choice reduces the computational complexity of generating the covariance matrix for large $n$.


\begin{Schunk}
\begin{Sinput}
R> M <- 120; n <- M^2
R> locs <- expand.grid(x=(0:(M-1))/(M-1), y=(0:(M-1))/(M-1))
R> D <- as.matrix(dist(locs))
R> theta_true <- c(1, 0.03)
R> cov_true <- cov.exponential(D, theta_true[1], theta_true[2])
R> set.seed(1234)
R> z <- t(chol(cov_true)) %*% rnorm(n)
\end{Sinput}
\end{Schunk}

\begin{Schunk}
\begin{Sinput}
R> # Negative log-likelihood
R> nll <- function(pars, prec='R-Double') {
+     nu <- 0.5      
+     a  <- exp(pars[1])             # a > 0
+     sigma2 <- exp(pars[2])         # sigma^2 > 0
+     
+     V <- cov.matern(D, nu, a, sigma2)
+     if(grepl('Single', prec)) diag(V) <- diag(V) + 1e-6
+     
+     use_mpcr <- grepl('MPCR', prec)
+     if(use_mpcr) {
+         # Precision
+         p <- if (prec == "MPCR-Single-CPU" || prec == "MPCR-Single-GPU")
+         "single" else "double"
+         # Hardware placement to expected values
+         op_place  <- if (prec == "MPCR-Single-CPU" ||
                       prec == "MPCR-Double-CPU")  "CPU" else "GPU"
+         V.mp <- as.MPCR(V, n, n, p, placement = op_place)
+         z.mp <- as.MPCR(z, n, 1, p, placement = op_place)
+         
+         U <- chol(V.mp)
+         L <- t(U)
+         
+         log_det <- 2 * log(diag(U))$Sum()
+         w <- forwardsolve(L, z.mp)
+         quad <- w$SquareSum()
+         
+     } else {
+         L <- t(chol(V))
+         log_det <- 2 * sum(log(diag(L)))
+         w <- forwardsolve(L, z)
+         quad <- sum(w^2)
+     }
+     
+     0.5*quad + 0.5*log_det + 0.5*n*log(2*pi)
+ }
\end{Sinput}
\end{Schunk}

\begin{Schunk}
\begin{Sinput}
R> run_mle <- function(label) {
+     cat(label, "optimization...\n")
+     t0 <- Sys.time()
+     fit <- optim(init, nll, prec=label, control=list(maxit=1000, trace=3))
+     time <- difftime(Sys.time(), t0, units="secs")
+     
+     theta_hat <- c(2*plogis(fit$par[1]), exp(fit$par[2]))
+     
+     cat("\n", label, "results:\n", sep="")
+     cat("  NLL:", fit$value, "\n")
+     cat("  Estimates:", 
+         sprintf("(%.7f, %.7f)", theta_hat[1], theta_hat[2]),
+         "\n")
+     cat("  Time:", round(time, 2), "secs\n\n")
+     
+     invisible(list(fit=fit, time=time, theta=theta_hat))
+ }
\end{Sinput}
\end{Schunk}

\begin{Schunk}
\begin{Sinput}
R> init <- c(1.5, -0.3)
R> res_R     <- run_mle("R-Double")
R> res_MP_D  <- run_mle("MPCR-Double-CPU")
R> res_MP_S  <- run_mle("MPCR-Single-CPU")
R> res_MP_D  <- run_mle("MPCR-Double-GPU")
R> res_MP_S  <- run_mle("MPCR-Single-GPU")
\end{Sinput}
\end{Schunk}

\begin{table}[h!]
\caption{MLE results under different precision types for $n = 14{,}400$ ($M = 120$). The column \texttt{nll} reports the Gaussian log-likelihood at the estimated parameters. {\color {red} $\nu$ was fixed at $0.5$ (true value) during the optimization process.}}
\centering
\scalebox{0.85}{
\begin{tabular}{c c c c c c c c c}
   \toprule
Precision & \texttt{nll} & Parameter Estimates & Num of Iterations & Execution Time \\[0.5ex] \hline
\proglang{R}-Double & $11607.29$ & $\hat{\boldsymbol{\theta}} = (0.9956609, 0.0297382, 0.5)^\top$ & 99 & 800.44 seconds \\
\pkg{MPCR}-Double-CPU & $11607.29$ & $\hat{\boldsymbol{\theta}} = (0.9956609, 0.0297382, 0.5)^\top$ & 99 & 757.46 seconds \\
\pkg{MPCR}-Single-CPU & $11607.29$ & $\hat{\boldsymbol{\theta}} = (0.9956609, 0.0297382, 0.5)^\top$ & 99 & 608.99 seconds \\
\pkg{MPCR}-Double-GPU & $11607.29$ & $\hat{\boldsymbol{\theta}} = (0.9956609, 0.0297382, 0.5)^\top$ & 99 & 677.41 seconds \\
\pkg{MPCR}-Single-GPU & $11607.29$ & $\hat{\boldsymbol{\theta}} = (0.9956609, 0.0297382, 0.5)^\top$ & 99 & 593.32 seconds \\
\hline
\end{tabular}
}
\label{tab:MPCR_JSS_application_mle_results}
\end{table}

{\color {red} Table~\ref{tab:MPCR_JSS_application_mle_results} presents the log-likelihood values, parameter estimates, number of iterations, and execution times for each precision setting. All configurations, including \proglang{R}-Double and the four \pkg{MPCR} variants, yield identical parameter estimates and log-likelihood values, thereby confirming numerical consistency across different precisions and hardware backends. Although the number of iterations remains constant, execution times differ across platforms. GPU implementations, particularly those using single-precision arithmetic, achieve the fastest runtimes. In this example, the estimated range parameter $a \approx 0.03$ suggests weakly correlated data and a well-conditioned covariance matrix, enabling single-precision to achieve accuracy comparable to double-precision. For more strongly correlated data (larger $a$), the covariance matrix becomes increasingly ill-conditioned, and single-precision arithmetic may lose accuracy, potentially requiring additional iterations and, in some cases, leading to a slightly biased or incorrect log-likelihood.}

\subsection{Principal Component Analysis (PCA)}

The \pkg{MPCR} package can also accelerate Principal Component Analysis (PCA), a widely used dimensionality-reduction technique in environmental and climate sciences, often referred to as Empirical Orthogonal Function (EOF) analysis. Given a spatio–temporal field $Z(\mathbf{s}, t)$, EOF analysis decomposes the data matrix into spatial modes and temporal scores via the singular value decomposition (SVD) $\mathbf{Z}=\mathbf{U}\mathbf{D}\mathbf{V}^\top$. For large space–time grids, the SVD can be computationally expensive, making precision selection particularly relevant.

We illustrate PCA using zonal wind measurements from the North Atlantic region (2010–2014). The data matrix is formed by stacking all spatial locations into columns, with each row corresponding to a time point. We compute the SVD using five precision settings: standard \proglang{R} double-precision, \pkg{MPCR}-Double, and \pkg{MPCR}-Single on both CPU and GPU.
\begin{Schunk}
\begin{Sinput}
R> library(ncdf4)
R> library(fields)
R> library(maps)
R> # Load climate data (download from Google Drive link)
R> link <- "b.e21.BHISTcmip6.f09_g17.LE2-1001.001.cam.h3.UBOT.2010010100-2014123100.nc"
R> nc <- nc_open(link)
R> lat <- ncvar_get(nc, "lat"); lon <- (ncvar_get(nc, "lon") + 180) %% 360 - 180
R> # Extract North Atlantic region  
R> lat_idx <- which(lat >= 7 & lat <= 85)
R> lon_idx <- which(lon >= -180 & lon <= -20)
R> ubot_raw <- ncvar_get(nc, "UBOT", start=c(lon_idx[1], lat_idx[1], 1),
+                       count=c(length(lon_idx), length(lat_idx), -1))
R> # Reshape to matrix (time x space)
R> valid <- which(!is.na(ubot_raw[,,1]))
R> ubot <- matrix(0, dim(ubot_raw)[3], length(valid))
R> for(i in 1:dim(ubot_raw)[3]) ubot[i,] <- ubot_raw[,,i][valid]
\end{Sinput}
\end{Schunk}
\begin{Schunk}
\begin{Sinput}
R> # Function to run PCA with different precisions
R> run_pca <- function(prec='R-Double') {
+   use_mpcr <- grepl('MPCR', prec)
+   t0 <- Sys.time()
+   
+   if(use_mpcr) {
+     # Precision
+     p <- if (prec == "MPCR-Single-CPU" || prec == "MPCR-Single-GPU")
+           "single" else "double"
+     # Hardware placement to expected values
+     op_place  <- if (prec == "MPCR-Single-CPU" ||
                       prec == "MPCR-Double-CPU")  "CPU" else "GPU"
+     X <- as.MPCR(ubot, nrow(ubot), ncol(ubot), p, placement = op_place)
+     s <- svd(X)
+     list(u=MPCR.ToNumericMatrix(s$u), v=MPCR.ToNumericMatrix(s$v),
+          d=MPCR.ToNumericVector(s$d), time=difftime(Sys.time(), t0, units="mins"))
+   } else {
+     s <- svd(ubot)
+     list(u=s$u, v=s$v, d=s$d, time=difftime(Sys.time(), t0, units="mins"))
+   }
+ }
\end{Sinput}
\end{Schunk}
\begin{Schunk}
\begin{Sinput}
R> # Run all precisions
R> res <- lapply(c('R-Double','MPCR-Double-CPU','MPCR-Single-CPU',
+ 'MPCR-Double-GPU','MPCR-Single-GPU'), run_pca)
R> names(res) <- c('R-Double','MPCR-Double-CPU','MPCR-Single-CPU',
+  'MPCR-Double-GPU','MPCR-Single-GPU')
R> 
R> # Visualizations
R> library(RColorBrewer)
R> # (1) Eigenvalues
R> pdf("pca_eigenvalues.pdf", width=16, height=4)
R> pct_var <- lapply(res, function(r) 100*r$d^2/sum(r$d^2))
R> matplot(1:20, sapply(pct_var, "[", 1:20), type="b", lty=1:3, 
+         pch=c(21,22,24), col=1:3, lwd=2, ylim=c(0,40),
+         xlab="Eigenvalue", ylab="Percentage of variance [%]",
+         main="Percent of Explained Variance")
R> legend("topright", names(res), col=1:3, lty=1:3, pch=c(21,22,24), lwd=2)
R> dev.off()
R> 
R> # (2) EOFs (3x3 grid)
R> pdf("pca_eofs.pdf", width=16, height=4)
R> par(mfrow=c(3,3), mar=c(4,4,3,2))
R> cols <- colorRampPalette(brewer.pal(9, "BrBG"))(100)
R> for(i in 1:3) for(nm in names(res)) {
+   Z <- matrix(NA, length(lon_idx), length(lat_idx))
+   Z[valid] <- if(is.matrix(res[[nm]]$v)) res[[nm]]$v[,i] else 
+               matrix(res[[nm]]$v, nrow=length(valid))[,i]
+   eof_label <- c("1st EOF", "2nd EOF", "3rd EOF")[i]
+   image.plot(lon[lon_idx], lat[lat_idx], Z, col=cols,
+              main=if(i==1) sprintf("%s (%.2f mins)", nm, res[[nm]]$time) else "",
+              xlab=if(i==3) "Longitude" else "", 
+              ylab=if(nm=="R-Double") "Latitude" else "",
+              cex.main=1.5, cex.lab=1.3)
+   if(nm=="R-Double") mtext(eof_label, side=2, line=5, cex=1.2, font=2)
+   lines(map("world", plot=FALSE)); box()
+ }
R> dev.off()
\end{Sinput}
\end{Schunk}

\begin{figure}[t!]
\centering
\includegraphics[width=\textwidth]{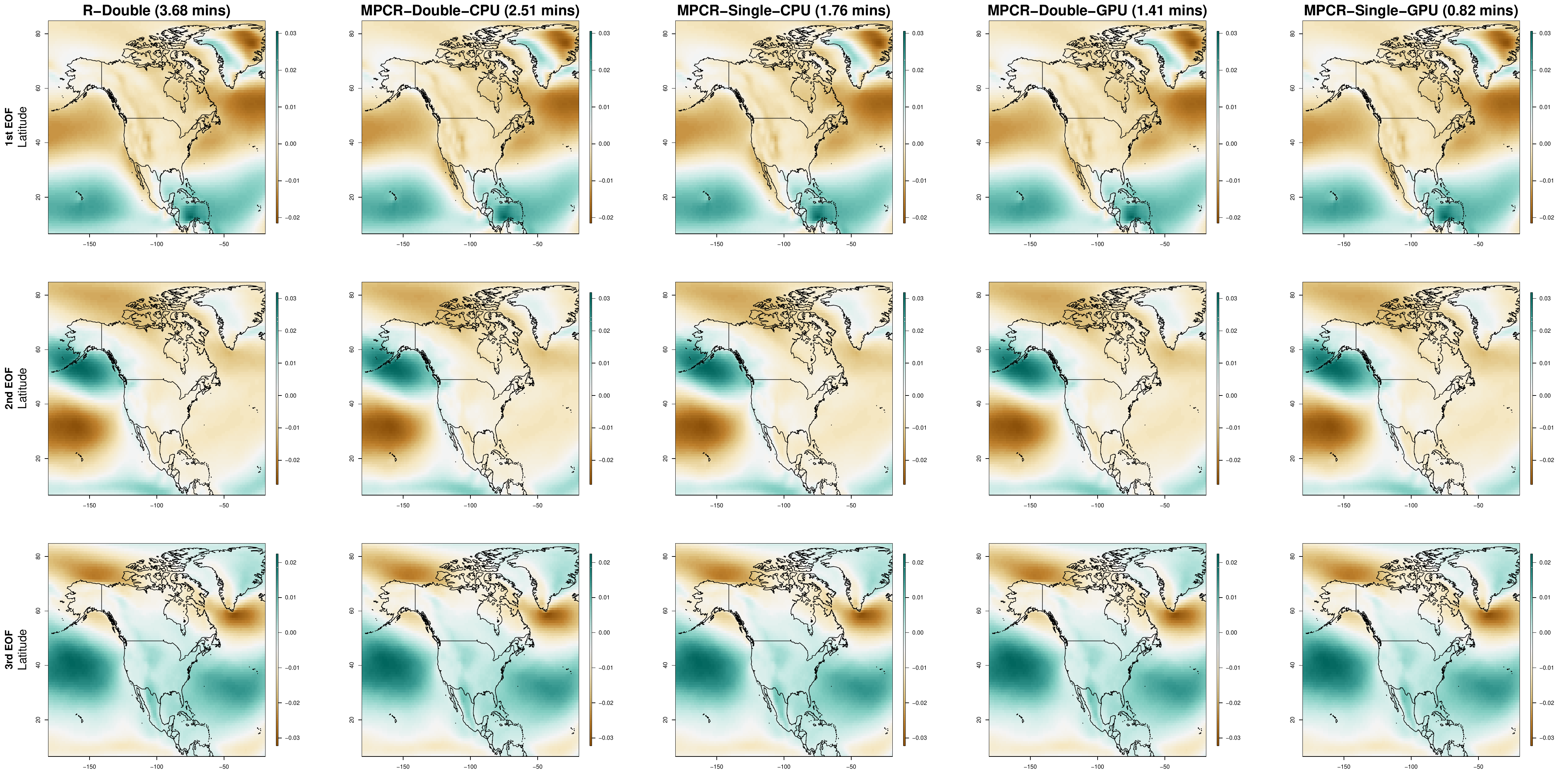}
\caption{First three EOFs of zonal wind measurements with execution times. All precision types produce visually identical spatial patterns. Some Panels (i.e., (2,4), (3,4), (2,5)) have been sign-aligned by multiplying the corresponding eigenvectors by $-1$. This adjustment reflects the intrinsic sign indeterminacy of the SVD, where each eigenvector is defined only up to a sign flip \citep{bro2008resolving}.}
\label{fig:MPCR_JSS_application_pca_result_eof}
\end{figure}

\begin{figure}[t!]
\centering
\includegraphics[width=0.6\textwidth]{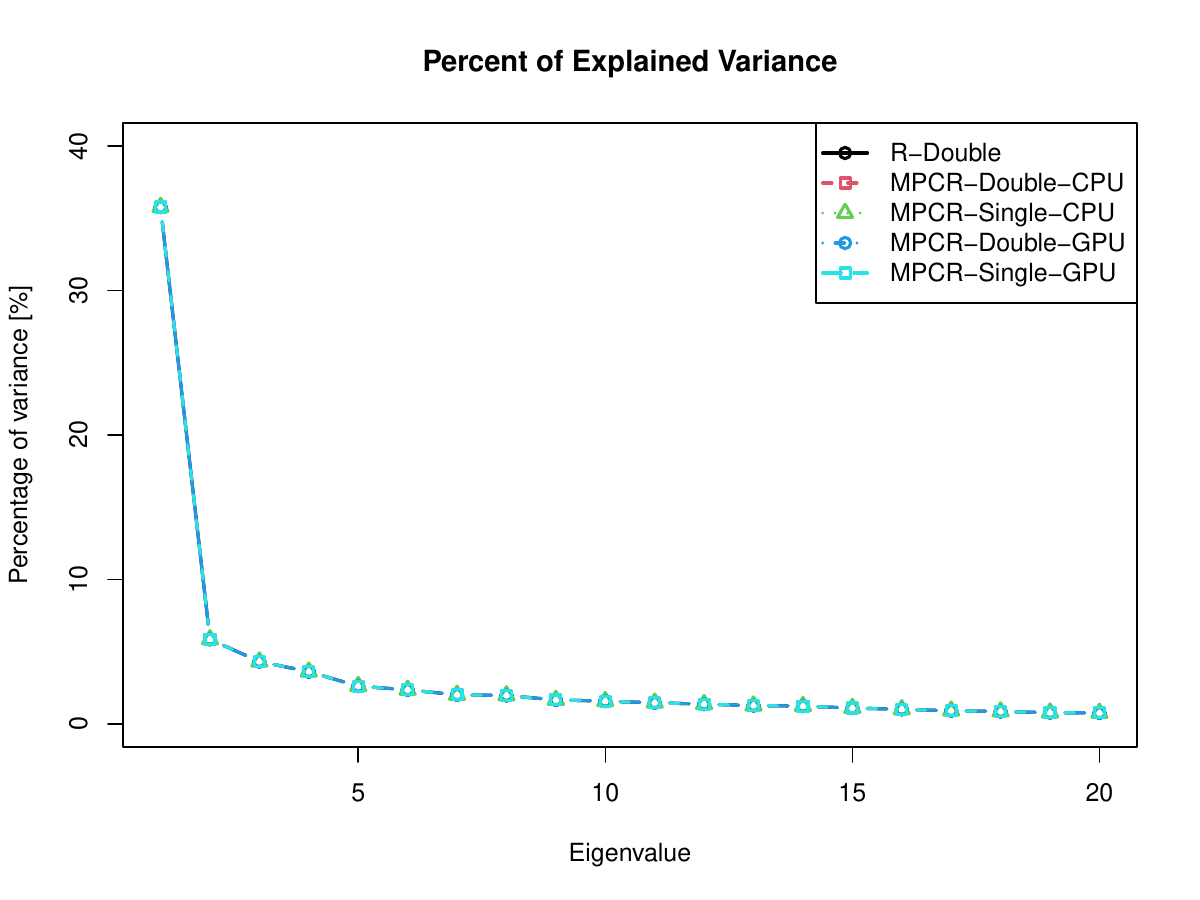}
\caption{Percentage of variance explained by the first 20 eigenvalues. The first three EOFs explain approximately 36\%, 6\%, and 4\% of total variance across all precision types.}
\label{fig:MPCR_JSS_application_pca_results_eigenvalue}
\end{figure}

Figure~\ref{fig:MPCR_JSS_application_pca_result_eof} displays the first three EOFs for each precision type. All methods produce indistinguishable spatial patterns—up to the sign indeterminacy inherent in any SVD/eigendecomposition, whose arbitrary resolution is implementation-dependent \citep{bro2008resolving}—confirming that reduced precision does not degrade the dominant variability modes. Similarly, the explained variance curves in Figure~\ref{fig:MPCR_JSS_application_pca_results_eigenvalue} show identical eigenvalue spectra across all precisions.

In terms of performance, GPU-based \pkg{MPCR} configurations outperform CPU and standard \proglang{R} implementations, with \pkg{MPCR}-Single-GPU achieving the fastest runtime (0.91 minutes), followed by \pkg{MPCR}-Double-GPU (1.58 minutes). Among CPU-based methods, \pkg{MPCR}-Single-CPU (2.65 minutes) and \pkg{MPCR}-Double-CPU (4.25 minutes) both outperform standard \proglang{R} (6.49 minutes). These results demonstrate that \pkg{MPCR} can substantially reduce computation time for high-dimensional PCA while preserving numerical fidelity. As with other examples, the key advantage is the ability to flexibly select lower precision depending on hardware and accuracy needs.

\subsection{Integrated Nested Multi-Precision Arithmetic Laplace Approximation}

Bayesian inference with latent Gaussian models often relies on the Integrated Nested Laplace Approximation (INLA) framework \citep{rue2009approximate, rue2017bayesian, bakka2018spatial}. INLA is highly efficient because it exploits sparsity in the precision matrix $\mathbf{Q}$, but challenges arise when models induce dense matrices or repeated Cholesky factorizations dominate runtime. In such cases, precision control becomes essential.

Using \pkg{MPCR}, we propose the Integrated Nested Multi-Precision Arithmetic Laplace Approximation, a direct analogue of INLA that stores and manipulates dense precision matrices in reduced precision (single). It follows the standard INLA workflow: Newton updates for the latent mode $\mathbf{x}_0(\alpha)$, Hessian formation, and Laplace evaluation, while delegating all expensive linear algebra operations to \pkg{MPCR} at the selected precision level.
\begin{Schunk}
\begin{Sinput}
R> library(MPCR)
R> library(Matrix)
R> # Global parameters
R> alpha_true <- 0.6; sigma_true <- 0.1; beta_true <- 10
\end{Sinput}
\end{Schunk}
\begin{Schunk}
\begin{Sinput}
R> # Helper to generate data for any n
R> gen_data <- function(n_size) {
+   locs <- cbind(seq(0, n_size, length.out=n_size), 0)
+   D <- as.matrix(dist(locs))
+   cov <- sigma_true * exp(-D/alpha_true)
+   set.seed(4); x <- drop(matrix(rnorm(n_size), 1, n_size) %*% t(chol(cov)))
+   p <- 1/(1 + exp(-beta_true*x))
+   list(D=D, y=rbinom(n_size, 1, p), n=n_size)
+ }
\end{Sinput}
\end{Schunk}
\begin{Schunk}
\begin{Sinput}
R> # INLA function
R> run_inla <- function(prec='R-Double', D, y, n) {
+   use_mpcr <- grepl('MPCR', prec)
+     # Precision
+     p <- if (prec == "MPCR-Single-CPU" || prec == "MPCR-Single-GPU") 
+           "single" else "double"
+     # Hardware placement to expected values
+     op_place  <- if (prec == "MPCR-Single-CPU" || prec == "MPCR-Double-CPU")
+                  "CPU" else "GPU"
+   alpha_vec <- seq(0.05, 0.95, length=21)
+   
+   calc_Q <- function(alpha) {
+     cov <- sigma_true * exp(-D/alpha)
+     if(use_mpcr) cov <- as.MPCR(cov, n, n, p, placement = op_place)
+     solve(cov)
+   }
+   
+   calc_x0 <- function(alpha, tol=1e-12) {
+     Q <- calc_Q(alpha)
+     x <- x0 <- rep(0, n)
+     repeat {
+       g1 <- beta_true*y - beta_true*exp(beta_true*x)/(1+exp(beta_true*x))
+       g2 <- -beta_true^2*exp(beta_true*x)/(1+exp(beta_true*x))^2
+       diag_g <- bandSparse(n=n, k=0, diagonals=list(g2))
+       if(use_mpcr) {
+         diag_g <- as.MPCR(as.matrix(diag_g), n, n, p, placement = op_place)
+         mode <- as.MPCR(g1-x0*g2, n, 1, p, placement = op_place)
+         x <- MPCR.ToNumericVector(solve(Q-diag_g) %*% mode)
+       } else {
+         x <- drop(solve(Q-diag_g) %*% (g1-x0*g2))
+       }
+       if(mean((x-x0)^2) < tol) break else x0 <- x
+     }
+     x
+   }
+   
+   calc_lpost <- function(alpha) {
+     x0 <- calc_x0(alpha)
+     Q <- calc_Q(alpha)
+     diag_term <- diag(beta_true^2*exp(beta_true*x0)/(1+exp(beta_true*x0))^2, n, n)
+     if(use_mpcr) diag_term <- as.MPCR(diag_term, n, n, p, placement = op_place)
+     H <- Q + diag_term
+     chol_Q <- chol(Q); chol_H <- chol(H)
+     
+     if(use_mpcr) {
+       logdet_Q <- log(diag(chol_Q))$Sum() * 2
+       logdet_H <- log(diag(chol_H))$Sum() * 2
+       x0_mpcr <- as.MPCR(x0, n, 1, p, placement = op_place)
+       quad <- (chol_Q %*% x0_mpcr)$SquareSum()
+     } else {
+       logdet_Q <- 2*sum(log(diag(chol_Q)))
+       logdet_H <- 2*sum(log(diag(chol_H)))
+       quad <- drop(t(x0) %*% Q %*% x0)
+     }
+     sum(beta_true*x0*y - log1p(exp(beta_true*x0))) + 
+       0.5*logdet_Q - 0.5*quad - 0.5*logdet_H
+   }
+   
+   t0 <- Sys.time()
+   lpost <- sapply(alpha_vec, calc_lpost)
+   time_elapsed <- as.numeric(difftime(Sys.time(), t0, units="secs"))
+   
+   lpost <- lpost - mean(lpost)
+   h <- alpha_vec[2] - alpha_vec[1]
+   w <- c(1, rep(c(4,2), (length(alpha_vec)-3)/2), 4, 1)
+   Z <- sum(w * exp(lpost)) * h/3
+   
+   list(alpha=alpha_vec, posterior=exp(lpost)/Z, time=time_elapsed)
+ }
\end{Sinput}
\end{Schunk}
\begin{Schunk}
\begin{Sinput}
R> # Compute for all sizes
R> sizes <- c(900, 2500)
R> precs <- c('R-Double', 'MPCR-Double-CPU', 'MPCR-Single-CPU', 
R>           'MPCR-Double-GPU', 'MPCR-Single-GPU')
R> timing_results <- matrix(NA, length(sizes), length(precs))
R> res <- NULL  # Will store n=2500 results for plotting
R> 
R> cat("Computing INLA for all sizes...\n")
R> for(i in seq_along(sizes)) {
+   cat("\nSize n =", sizes[i], "\n")
+   d <- gen_data(sizes[i])
+   
+   for(j in seq_along(precs)) {
+     cat("  ", precs[j], "...")
+     result <- run_inla(precs[j], d$D, d$y, d$n)
+     timing_results[i,j] <- result$time
+     
+     # Save n=2500 results for posterior plot
+     if(sizes[i] == 2500 && is.null(res)) {
+       res <- list()
+     }
+     if(sizes[i] == 2500) {
+       res[[precs[j]]] <- result
+     }
+     
+     cat(" ", round(timing_results[i,j], 2), "secs\n")
+   }
+ }
R> 
R> colnames(timing_results) <- precs
R> rownames(timing_results) <- paste0("n=", sizes)
R> cat("\nTiming Results (seconds):\n")
R> print(timing_results)
\end{Sinput}
\end{Schunk}
\begin{Schunk}
\begin{Sinput}
R> # Posterior plot for n=2500
R> pdf("inla_posterior.pdf", width=10, height=8)
R> cols <- c("black", "red", "blue"); ltys <- c(1, 2, 4)
R> plot(res[[1]]$alpha, res[[1]]$posterior, type="l", lwd=2.5, col=cols[1], 
+      lty=ltys[1], xlab=expression(alpha), ylab="Posterior", ylim=c(0,4.5),
+      main=expression(paste("Posterior Distribution of ", alpha, " (n=2500)")))
R> for(i in 2:3) {
+   lines(res[[i]]$alpha, res[[i]]$posterior, lwd=2.5, col=cols[i],
+   lty=ltys[i])
+ }
R> for(i in 1:3) {
+   idx <- seq(1, 21, by=3)
+   points(res[[i]]$alpha[idx], res[[i]]$posterior[idx], pch=c(16,17,15)[i], 
+   col=cols[i], cex=0.8)}
R> legend("topright", names(res), col=cols, lwd=2.5, lty=ltys, pch=c(16,17,15), 
+   pt.cex=0.8)
R> abline(v=alpha_true, lty=3, col="gray", lwd=1.5)
R> dev.off()
\end{Sinput}
\end{Schunk}
\begin{Schunk}
\begin{Sinput}
R> # Timing plot
R> pdf("inla_timing.pdf", width=8, height=6)
R> matplot(sizes, timing_results, type="b", lty=1:3, pch=c(21,22,24), 
+         col=1:3, lwd=2, log="y",
+         xlab="Problem Size (n)", ylab="Time (seconds, log scale)",
+         main="INLA Computation Time vs Problem Size")
R> legend("topleft", precs, col=1:3, lty=1:3, pch=c(21,22,24), lwd=2)
R> grid(TRUE, lty=2, col="gray90")
R> dev.off()
\end{Sinput}
\end{Schunk}

\begin{figure*}[t!]
    \centering
    \begin{subfigure}[t]{0.5\textwidth}
        \centering
        \includegraphics[width=\textwidth]{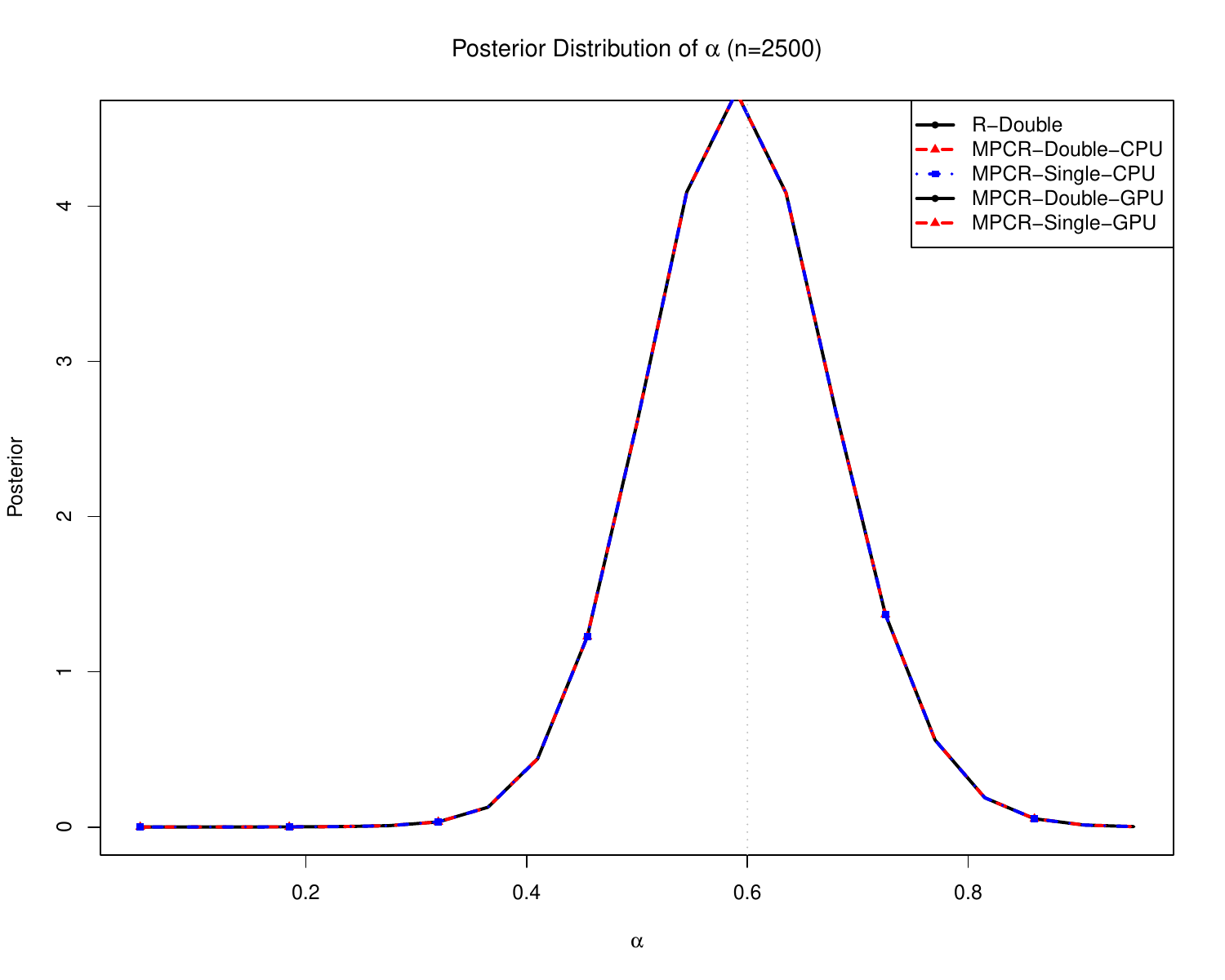}
        \caption{Posterior distribution of $\alpha$}
        \label{fig:MPCR_JSS_application_inla_result_posterior}
    \end{subfigure}%
    ~ 
    \begin{subfigure}[t]{0.55\textwidth}
        \centering
        \includegraphics[width=\textwidth]{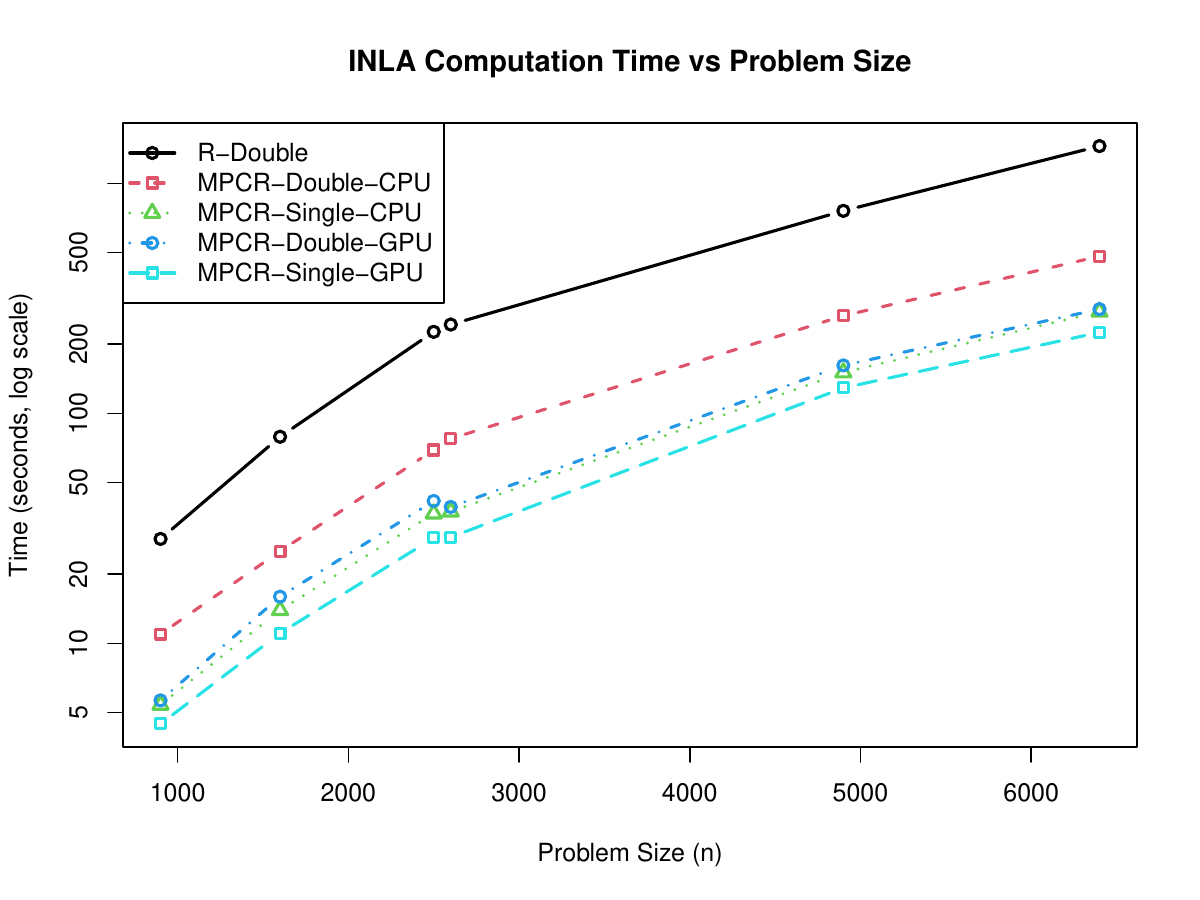}
        \caption{Execution Time}
        \label{fig:MPCR_JSS_application_inla_result_time}
    \end{subfigure}
    \caption{Results of the INLA experiments: (a) Approximation of the posterior distribution of $\alpha$ for $n = 2{,}500$. The bold vertical line marks the true value of $\alpha = 0.6$. (b)~Execution time of computing the posterior distribution for various $n$ under different precisions.}
    \label{fig:INSHALA_results}
\end{figure*}

All five precision settings yield identical posterior distributions for the hyperparameter $\alpha$. For $n=2{,}500$, the posterior curves coincide across all precision types, as shown in Figure~\ref{fig:INSHALA_results}(a). 

{\color{red} The primary difference lies in computational efficiency. Figure~\ref{fig:INSHALA_results}(b) shows that \pkg{MPCR}-Double reduces runtime compared with standard \proglang{R}, while \pkg{MPCR}-Single delivers the largest speedups, particularly for larger $n$. Across problem sizes, \pkg{MPCR}-Double-CPU achieves roughly $2.6\times$–$3.3\times$ speedup, whereas \pkg{MPCR}-Single-CPU reaches about $5.0\times$–$6.5\times$. On accelerators, \pkg{MPCR}-Double-GPU provides around $4.7\times$–$6.2\times$ speedup, and \pkg{MPCR}-Single-GPU delivers the highest gains, ranging from approximately $5.9\times$ to $8.4\times$. These improvements stem solely from using lower numerical precision, without any changes to the statistical model or approximation scheme.}

\section{Discussion and Future Work} \label{sec:discuss}

Multi-precision computation can substantially accelerate the processing of large-scale scientific problems. This paper introduces \pkg{MPCR}, a package developed to enable multi-precision computation within the \proglang{R} environment. \pkg{MPCR} can be linked to optimized BLAS/LAPACK libraries, including Intel MKL, OpenBLAS, and cuBLAS, to achieve optimal hardware performance. The functionality of the package is demonstrated in \proglang{R} through several examples, illustrating how it improves execution efficiency by reducing computational precision while maintaining double-precision accuracy. \pkg{MPCR} supports three levels of precision for R operations: double-precision, single-precision, and half-precision (for matrix-matrix multiplications only). Additionally, \pkg{MPCR} provides a robust framework that supports higher precision, such as 128-bit arithmetic, contingent on compatibility with the underlying BLAS/LAPACK libraries and the hardware architecture. Its seamless integration with optimized BLAS/LAPACK libraries establishes \pkg{MPCR} as an effective alternative to standard \proglang{R} operations, especially when \proglang{R} is linked to less-optimized libraries such as RBLAS.

In future work, we plan to extend \pkg{MPCR} to support mixed-precision matrix operations, allowing the same matrix or vector to be represented at multiple precisions. This will enable more efficient linear algebra algorithms, such as tiled approaches that exploit parallel execution across different precisions (e.g., 16-, 32-, and 64-bit). This extension will utilize OpenMP for parallelization on multicore systems and incorporate GPU acceleration. These enhancements are expected to allow \proglang{R} users to perform linear algebra operations more efficiently, providing substantial speedups compared to computations performed exclusively at high precision.







\section*{Computational details}

All experiments in this paper were conducted under the following software and hardware environment:
\begin{verbatim}
R version 4.4.2
Platform: x86_64-pc-linux-gnu
Running under: Ubuntu 22.04.5 LTS
BLAS/LAPACK: MKL
\end{verbatim}

The analyses were performed using the \pkg{MPCR} package (version~2.0.0). \proglang{R}, \pkg{MPCR}, and all other
packages used in this paper are available from the Comprehensive \proglang{R}
Archive Network (CRAN) at \url{https://CRAN.R-project.org/}.

\section*{Acknowledgment}

\begin{leftbar}
This study was funded by King Abdullah University of Science and Technology (KAUST), Saudi Arabia. We thank the BrightSkies team for their technical support in developing the \pkg{MPCR} package, and we are especially grateful to Omar Marzouk, Merna Moawad, and Ali Hakam (BrightSkies Inc.) for their invaluable contributions.
\end{leftbar}


\bibliography{refs}

\end{document}